\documentclass[journal]{IEEEtran}
\usepackage{latexsym}
\usepackage{graphicx}
\usepackage{amsfonts,amssymb,amsmath}
\usepackage{hyperref}
\usepackage{xcolor}
\usepackage{subcaption}
\usepackage{epstopdf}
\usepackage{color}
\usepackage{tikz}
\usepackage{pgfplots} 
\usetikzlibrary{patterns}
\def\BibTeX{{\rm B\kern-.05em{\sc i\kern-.025em b}\kern-.08em T\kern-.1667em\lower.7ex\hbox{E}\kern-.125emX}}
\usepackage{textcomp}
\def\BibTeX{{\rm B\kern-.05em{\sc i\kern-.025em b}\kern-.08em
		T\kern-.1667em\lower.7ex\hbox{E}\kern-.125emX}}

\usepackage[noadjust]{cite}

\DeclareMathOperator{\K}{K}
\DeclareMathOperator{\G}{G}

\begin{document}

\title{On the physical layer security capabilities of reconfigurable intelligent surface empowered wireless systems}

\author{Alexandros--Apostolos A. Boulogeorgos,~\IEEEmembership{Senior Member, IEEE},  Angeliki Alexiou,~\IEEEmembership{Member, IEEE},  and \\ Angelos Michalas,~\IEEEmembership{Member, IEEE}

\thanks{This work was supported from the European Commission's Horizon 2020 research and innovation programme under grant agreement  No. 871464 (ARIADNE).}
\thanks{A.-A. A. Boulogeorgos and A. Michalas are with the Department of Electrical and Computer Engineering, University of Western Macedonia, Kozani, , Greece (e-mails: aboulogeorgos@uowm.gr, amichalas@uowm.gr)}
\thanks{A. Alexiou is with the Department of Digital Systems,
	University of Piraeus
	Piraeus 18534 Greece (e-mail:  alexiou@unipi.gr).}
 }

\maketitle
\begin{abstract}
In this paper, we investigate the physical layer security capabilities of reconfigurable intelligent surface (RIS) empowered wireless systems. In more detail, we consider a general system model, in which the links between the transmitter (TX) and the RIS as well as the links between the RIS and the legitimate receiver are modeled as mixture Gamma (MG) random variables (RVs). Moreover, the link between the TX and eavesdropper is also modeled as a MG RV.	Building upon this system model, we derive the probability of zero-secrecy capacity as well as the probability of information leakage. Finally, we extract the average secrecy rate for both cases of TX having full and partial channel state information knowledge. 
\end{abstract}

\begin{IEEEkeywords}
 Average secrecy rate, Mixture Gamma, Physical layer securtiy, Performance Analysis, Probability of information leakage, Probability of zero-secrecy rate, Reconfigurable intelligent surfaces. 
\end{IEEEkeywords}


\section{INTRODUCTION}

\IEEEPARstart{A}{s} the wireless world shifts its attention towards higher frequency bands, such as the millimeter wave and the terahertz (THz) band, the nature of the wireless links changes from omni- or semi-omni- to pencil-directional~\cite{Boulogeorgos2018a,WP:Wireless_Thz_system_architecture_for_networks_beyond_5G,Boulogeorgos2019}. This fundamental change provides the opportunity to design and deploy lightweight physical layer security (PLS) schemes~\cite{Boulogeorgos2021,Tsiftis2022,Yang2022}. PLS ensure information-theoretic security without decreasing either the energy or the data efficiency of the wireless system~\cite{Wu2018,Wang2019,8710252}. In other words, they can be seen as an additional security layer that complements other methods and enhance the trustworthiness of next-generation radio access networks~\cite{Mitev2023}. 

Motivated by this, several researchers studied the performance of PLS schemes in high-frequency systems~\cite{Ju2017,Vuppala2018,Wu2020,Zhu2017,Wang2022,Huang2020,Fang2022,Mei2021,Ma2018}.  For instance, in~\cite{Ju2017}, the authors investigated the secrecy outage performance of a millimeter wave (mmW) wireless system, in which the legitimate transmitter (TX) employs either full or partial maximum ratio transmitting beamforming, while both the legitimate receiver (RX) and the eavesdropper use omni-directional antennas. Likewise, a performance analysis of mmW-overlaid microwave cellular networks, from a security perspective, which returns the mathematical framework for the derivation of the secrecy outage probability, and the achievable average secrecy rate, was presented in~\cite{Vuppala2018}. 
The authors of~\cite{Wu2020} reported an asymptotic closed-form expression for the secrecy rate of lens antenna array wireless mmW systems. The  impact of mmW channel particularities including random blockages and antenna gains on the secrecy performance was documented in~\cite{Zhu2017}. Moreover, in~\cite{Wang2022}, the authors investigated the potential of PLS in mmW multiple-input–multiple-output  systems in the presence and absence of artificial noise (AN), and reported a lower bound of the secrecy outage probability (SOP) for the case of AN as well as a exact-form of the SOP without AN. In~\cite{Huang2020}, the mathematical framework for the SOP for mmW non-orthogonal multiple access networks, in which legitimate users and eavesdroppers are randomly distributed, was presented.

In~\cite{Fang2022}, the authors studied the resilience against eavesdropping in long-range directional line-of-sight (LoS) outdoor links. In this type of links, the characteristic dimension of water molecules is comparable to the wavelength; thus, they act as scatterers. As a result, eavesdroppers that are not in LoS can capture the legitimates TXs messages. Experimental results revealed the feasibility and efficiency of this type of schemes. For the same system model, the authors of~\cite{Mei2021} provided a closed-form expression for the secrecy outage probability assuming that the wireless THz system suffers from atmospheric turbulence. The analysis highlighted that even under atmospheric turbulence, PLS can boost the security performance of outdoor THz wireless systems. The authors of~\cite{Mei2021} quantified the eavesdropping risk on THz wireless systems that experience atmospheric turbulence. Finally, in~\cite{Ma2018}, the authors explained that one of the key factor for successful eavesdropping in THz wireless systems is reflections due to~blockage. 

To counterbalance this, the use of reconfigurable intelligent surfaces (RISs) in order to manipulate the wireless environment and avoid blockers through the creation of alternative paths was investigated~\cite{Khoshafa2021,Ren2023,Yang2020,Shi2023}. In particular, in~\cite{Khoshafa2021}, the authors studied the secrecy performance in terms of secrecy outage probability of a RIS-aided PLS enhancement in device-to-device underlay wireless systems. The authors of~\cite{Ren2023} reported approximations for the sum achievable security data rate  for RIS-aided systems with statistical channel state information, assuming a Rician distribution for the basestation-RIS and RIS legitimate user and RIS-eavesdropper channel coefficients.  The SOP of a RIS-assisted wireless systems in which all the channels are modeled as Rayleigh distributed random variables (RVs) is reported in~\cite{Yang2020}. In~\cite{Shi2023}, the authors reported a tight bound for the SOP of a RIS-assisted wireless systems, in which the eavesdropper establishes a direct link with the TX. 

In all the aforementioned contributions, the eavesdropper establishes a link through the RIS. However, another very interesting scenario exists, in which the eavesdropper cannot receive the reflected by the RIS signal because either the reflection direction is blocked or the eavesdropper is located outside the RIS main lobe. This scenario was investigated in~\cite{Gu2023}, where the authors reported closed-form expressions for the probability of the nonzero secrecy capacity and the ergodic secrecy capacity, assuming that both the legitimate and eavesdropping channels can be modeled as Rayleigh RVs. This analysis provided some useful insights for the design of RIS-empowered PLS schemes, however, as the wireless world moves towards higher frequencies, the need of considering more general fading channels that capture the characteristics of directional links becomes more imperative. Motivated by this as well as that to the best of our knowledge, no general theoretical framework for the quantification of the secrecy performance of RIS-empowered wireless systems in which the eavesdropper cannot receive signal from the RIS exist, in this paper, we present the following contribution:
\begin{itemize}
    \item We report a general system model in which the legitimate RX captures the transmitted signal through an RIS, while the eavesdropper uses a direct link. All the established links are modeled as mixture Gamma (MG) RVs. Note that MG  is a general distribution that can be simplified in or accurately approximate a number of usually employed distributions including Rayleigh, Rice, Nakagami-$m$, $\alpha-\mu$, $\kappa-\eta$, etc. As a consequence, it can be used for modeling a number of low- to high-frequency wireless systems, including mmW~\cite{8737950,9051684,9464739,8839069,9419042} and THz~\cite{9887787,Papasotiriou2021,Papasotiriou2023}.
    \item Building upon the system model, we statistically characterize in terms of probability density function (PDF) and cumulative distribution function (CDF) of the signal-to-noise-ratio (SNR) at the legitimate RX and the~eavesdropper.
    \item We present novel closed-form expressions for the probability of zero secrecy capacity as well as the probability of information leakage. Finally, the average secrecy rate is extracted for both cases of full and partial channel state information (CSI) knowledge. Note that the term ``partial CSI'' is used to denote the realistic scenario in which the TX has full CSI knowledge of the legitimate link and no CSI knowledge of the eavesdropping link. 
\end{itemize}

The organization of the rest of this contribution is as follows: In Section~\ref{S:SM}, the system and channel model as well as the statistical characterization of the TX-RIS-legitimate RX channel are document. Section~\ref{S:SPA} focuses on presenting the secrecy performance analysis. Numerical results and insightful discussions are presented in Section~\ref{S:NR}. Finally, a summation of the paper and its key messages are reported in Section~\ref{S:Con}. 

\textit{Notations}: In what follows, $\sum_{m=1}^M x_m$ returns the sum of all the $x_m$ with $m\in[1, M]$. The absolute values is donated as $|\cdot|$, while $x^n$ stands for the $x$ in the power of $n$. $\Pr\left(\mathcal{A}\right)$ represents the probability of event $\mathcal{A}$. Moreover, $\frac{\rm{d}f(x)}{\rm{dx}}$ and $\int_{a}^{b}f(x)\,\rm{dx}$ respectively return the derivative and the integral of $f(x)$ from $a$ to $b$, respectively. The expected value of the RV $t$ is returned by $\mathbb{E}[t]$. The exponential function is represented by $\exp(\cdot)$ and the logarithmic function with base $2$ is $\log_2(\cdot)$. The operator $\max\left(a,b\right)$ returns the maximum value between $a$ and $b$. The lower incomplete Gamma and the Gamma functions are respectively denoted as $\gamma(\cdot, \cdot)$ and $\Gamma(\cdot)$, while the modified Bessel function of the second kind and order $n$ is written as $\rm{K}_n(\cdot)$. The generalized hypergeometric function is represented by $\,_pF_q\left(a;b;z\right)$. Finally,  $G_{m,n}^{p,q}\left(\cdot\left|\begin{array}{c} \cdots\\ \cdots \end{array}\right.\right)$ returns the Meijer-G~function.  

\section{System \& channel models}\label{S:SM}

This section aims to present the system and channel models. In more detail, the system model and the basic assumptions are reported in Section~\ref{SS:SM}, while the channel models and their statistical characterization are documented in Section~\ref{SS:CM}.

\subsection{System model}\label{SS:SM}

\begin{figure}
	\centering
	\scalebox{0.5}{\input{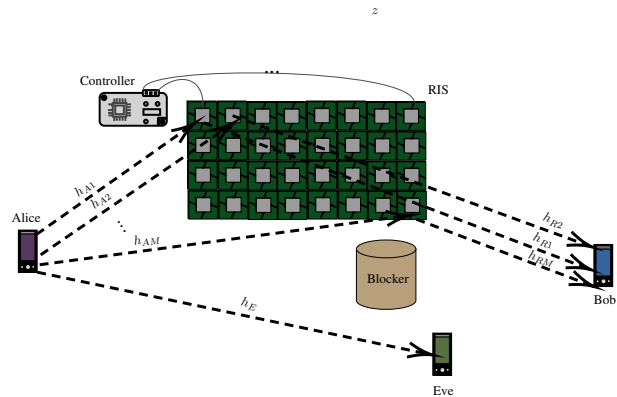}}
	\caption{System model.}
	\label{Fig:SM}
\end{figure}

As depicted in Fig.~\ref{Fig:SM}, we consider a scenario in which the TX, Alice, communicates with the legitimate RX, Bob, through an RIS in the presence of an eavesdropper, Eve. The RIS consists of $M$ meta-atoms (MAs) that are capable of conducting continuous phase shift (PS) and are coordinated by a micro-controller. Note that although continuous PS is very difficult or even impossible to be achieved in practice, there are several published contribution that prove that even with discrete PS MAs with beyond a specific number of states, an RIS-empowered wireless system with discrete PS can achieve similar performance with the corresponding one with continuous PS~\cite{Xu2021}. Moreover, the direct link between Alice and Bob is assumed to be blocked. Thus, the baseband equivalent received signal at Bob can be expressed~as in~\cite{9095301}
\begin{align}
 y_B = A_r \, s + n_B,
 \label{Eq:y_B}
\end{align}
where $s$ is Alice transmitted symbol, while $n_B$ stands for the additive white Gaussian noise and is modeled as a zero-mean complex Gaussian process of variance $\sigma_B^2$. In~\eqref{Eq:y_B}, $A_r$ represents the Alice-RIS-Bob baseband equivalent channel coefficient and can be analyzed~as
\begin{align}
    A_r = \sum_{m=1}^{M} h_{Am}\,r_m\,h_{Rm},
    \label{Eq:A_r}
\end{align}
with $h_{Am}$ and $h_{Rm}$ respectively being the Alice-$m$-th MA and $m-$th MA-Bob channel coefficients. Moreover, in~\eqref{Eq:A_r}, $r_m$ denotes the the $m-$th MA response and can be written~as
\begin{align}
    r_m = \left|r_m\right|\,\exp\left(j\,\phi_{m}\right),
    \label{Eq:r_m}
\end{align}
where $\left|r_m\right|$ and $\phi_{m}$ respectively denote the amplitude and phase of the response of the $m-$th MA. Assuming that $\left|r_m\right|\approx 1$, which, according to~\cite{PhysRevB.94.075142}, is a realistic assumption,~\eqref{Eq:r_m} can be rewritten~as
\begin{align}
    r_m \approx \exp\left(j\,\phi_{m}\right),
    \label{Eq:r_m}
\end{align}
Each RIS MA is assumed to have selected the optimal phase shift, i.e.,
\begin{align}
    \phi_m = - \theta_{Am} - \theta_{Bm},
\end{align}
where $\theta_{Am}$ and $\theta_{Bm}$ are respectively the phases of $h_{Am}$ and $h_{Bm}$. As a result,~\eqref{Eq:A_r} can be expressed~as
\begin{align}
    A_r = \sum_{m=1}^{M} \left|h_{Am}\right|\,\left|h_{Rm}\right|.
    \label{Eq:A_r_final}
\end{align}

Since the RIS steers the incident beam towards Bob, only a small portion of the incident signal from the RIS can be captured by Eve. By assuming that the RIS consists of a relative large number of MAs, which is a realistic assumption in high-frequency systems, the captured signal by Eve from RIS is considered to be below Eve's receiver noise threshold. As a consequence and by assuming the the direct Alice-Eve link is not blocked, the baseband equivalent received signal at Eve can be obtained~as
\begin{align}
    y_{E} = h_{E}\,s + n_{E},
    \label{Eq:y_E}
\end{align}
where $h_{E}$ is the Alice-Eve channel coefficient and $n_{E}$ denotes the additive noise that is modeled as a zero-mean complex Gaussian process of variance $\sigma_{E}^2$.

\subsection{Channel model}\label{SS:CM}

To create a general framework for the assessment of the performance of RIS-empowered wireless systems, we model the Alice-$m$-th MA, and $m-$MA-Bob,   channel coefficient envelops as independent MG RVs with PDF and CDF that can be respectively obtained~as
\begin{align}
f_{h_{\mathcal{X}m}}(x) = \sum_{n=1}^{N_{\mathcal{X}}} 2\, a_{n}^{\mathcal{X}}\,x^{2 b_{n}^{\mathcal{X}}-1}\exp\left(-c_{\mathcal{X}}\,x^2\right) 
\end{align}
and
\begin{align}
F_{h_{\mathcal{X}m}}(x) = \sum_{n=1}^{N_{\mathcal{X}}}  a_{n}^{\mathcal{X}}\, c_{\mathcal{X}}^{-b_{n}^{\mathcal{X}}}\,\gamma\left(b_{n}^{\mathcal{X}}, x^{2}\right),
\end{align}
where $m\in[1, M]$ and $\mathcal{X}\in\{A, R\}$. Likewise, $a_{n}^{\mathcal{X}}$, $b_{n}^{\mathcal{X}}$, and $c_{\mathcal{X}}$ are parameters of the $n-$th term of the MG distribution. Notice that MG can accurately model a number of well-known fading conditions, such as Rayleigh, Rice, Nakagami-$m$, Gamma, $\alpha-\mu$, etc.

As a sum of products of independent MG RVs, the PDF and CDF of $A_r$ can be respectively obtained as in~\cite{7460203} 
\begin{align}
    f_{A_r}\left(x\right) = \frac{4 \Xi^{k_A + m_A}}{\Gamma\left(k_A\right)\Gamma\left(m_A\right)} x^{k_A+m_A-1} \mathrm{K}_{k_A-m_A}\left(2\Xi x\right)
		\label{Eq:f_A}
	\end{align}
	and  \vspace{-0.3cm}
\begin{align}
		F_{A_r}(x) = \frac{1}{\Gamma\left(k_A\right)\Gamma\left(m_A\right)} \mathrm{G}_{1, 3}^{2, 1}\left(\Xi^2 x^2\left|\begin{array}{c} 1\\ k_A, m_A, 0 \end{array}\right.\right),
		\label{Eq:F_A}
\end{align}
with 
\begin{align}
		k_A = - \frac{b_A}{2 a_A} + \frac{\sqrt{b_A^2 - 4 a_A c_A}}{2 a_A},
		\label{Eq:k_A}
\end{align}  
\begin{align}
		m_A = - \frac{b_A}{2 a_A} - \frac{\sqrt{b_A^2 - 4 a_A c_A}}{2 a_A}
\end{align}
and  
\begin{align}
		\Xi = \sqrt{\frac{k_A m_A}{\Omega_A}}.
		\label{Eq:Xi_A}
\end{align}
In~\eqref{Eq:k_A}--\eqref{Eq:Xi_A},
\begin{align}
	a_A \hspace{-0.15cm}&=\hspace{-0.15cm} \mu_{A}\left(6\right) \mu_{A}\left(2\right) + \left(\mu_{A}\left(2\right)\right)^2 \mu_{A}\left(4\right) - 2\left(\mu_{A}\left(4\right)\right)^2,
	\\
	b_A\hspace{-0.15cm}&=\hspace{-0.15cm} \mu_{A}\left(6\right) \mu_{A}\left(2\right)- 4 \left(\mu_{A}\left(4\right)\right)^2 + 3 \left(\mu_{A}\left(2\right)\right)^2 \mu_{A}\left(4\right),
\\
	c_A &= 2 \left(\mu_{A}\left(2\right)\right)^2 \mu_{A}\left(4\right)
\end{align}
and \vspace{-0.3cm}
\begin{align}
	\Omega_A = \mu_A\left(2\right),
\end{align}
where \vspace{-0.3cm}
\begin{align}
	\mu_{A}\left(l\right) = \sum_{l_1=0}^{l}&\sum_{l_2=0}^{l_1}\cdots\sum_{l_{N-1}=0}^{l_{N-2}}
	\left(\begin{array}{c}l\\l_1\end{array}\right) \left(\begin{array}{c}l_1\\l_2\end{array}\right)
	\cdots 
	\left(\begin{array}{c}l_{N-2}\\l_{N-1}\end{array}\right)
	\nonumber \\ 
	& \hspace{-0.7cm} \times \mu_{\chi_1}\left(l-l_1\right) \mu_{\chi_2}\left(l_1-l_2\right) 
	\cdots \mu_{\chi_{N-1}}\left(l_{N-1}\right)
	\label{Eq:mu_A}
\end{align}
and 
\begin{align}
	\mu_{\chi_i}(l) = \sum_{m=1}^{M}&\sum_{k=1}^{K}
	a_m^{(1)} a_k^{(2)} \left(\frac{c_1}{c_2}\right)^{-\frac{b_m^{(1)}-b_k^{(2)}}{2}} \left(c_1 c_2\right)^{-\frac{b_m^{(1)}+b_k^{(2)}+n}{2}}
	\nonumber \\ & \times
	\Gamma\left(b_{m}^{(1)}+\frac{n}{2}\right) \Gamma\left(b_{k}^{(2)}+\frac{n}{2}\right)
	\label{Eq:mu_x}
\end{align}

Similarly, the Alice-Eve channel envelop is assumed to follow MG distribution with PDF and CDF that can be respectively expressed~as
\begin{align}
f_{h_{E}}(x) = \sum_{n=1}^{N_{E}} 2\, a_{n}^{E}\,x^{2 b_{n}^{E}-1}\exp\left(-c_{E}\,x^2\right) 
\label{Eq:f_h_E}
\end{align}
and
\begin{align}
F_{h_{E}}(x) = \sum_{n=1}^{N_{E}}  a_{n}^{E}\, c_{E}^{-b_{n}^{E}}\,\gamma\left(b_{n}^{E}, x^{2}\right),
\label{Eq:F_h_E}
\end{align}
where  $a_{n}^{E}$, $b_{n}^{E}$, and $c_{E}$ are parameters of the $n-$th term of the MG distribution that consists of $N_{E}$ terms.

\section{Secrecy performance analysis}\label{S:SPA}

This section focuses on quantifying the security performance of the RIS-empowered PLS wireless systems. In this direction, in Section~\ref{SS:SNR} closed form expressions for the SNR at Bob and Eve ends are presented. Moreover, in Section~\ref{SS:Full_CSI}, we assess the probability of zero-secrecy capacity and the average secrecy rate for the case of full CSI knowledge, while, in Section~\ref{SS:PCSI}, the probability of information leakage as well as the average secrecy rate for the case of partial CSI knowledge are reported.

\subsection{SNR statistical characterization}\label{SS:SNR}

Let $\gamma_{B}$ be the SNR at Bob's end, which, based on~\eqref{Eq:y_B}, can be obtained~as 
\begin{align}
    \gamma_{B} = \frac{A^{2}\,P_B}{\sigma_B^2},
    \label{Eq:gamma_B}
\end{align}
where $P_B$ stands for the Alice transmission power multiplied by the Alice-RIS-Bob link geometric gain. Moreover, Let $\gamma_{E}$ denote the SNR at Eve's end and, according to~\eqref{Eq:y_E} can be expressed~as
\begin{align}
    \gamma_{E} = \frac{\left|h_{E}\right|^{2}\,P_E}{\sigma_{E}^{2}},
    \label{Eq:gamma_E}
\end{align}
with $P_{E}$ representing the the Alice transmission power multiplied by the Alice-Eve link geometric gain. 
The following lemmas return the CDFs and PDFs of $\gamma_{B}$ and $\gamma_{E}$.

\textbf{Lemma 1:} The CDF and PDF of $\gamma_B$ can be respectively obtained~as
\begin{align}
    F_{\gamma_{B}}\left(x\right) =\frac{1}{\Gamma\left(k_A\right)\Gamma\left(m_A\right)} \mathrm{G}_{1, 3}^{2, 1}\left(\Xi^2 \frac{x\,\sigma_B^2}{P_B}\left|\begin{array}{c} 1\\ k_A, m_A, 0 \end{array}\right.\right)
    \label{Eq:F_gamma_B}
\end{align}
and
\begin{align}
    &f_{\gamma_{B}}\left(x\right) = \frac{1}{\Gamma\left(k_A\right)\Gamma\left(m_A\right)} \left(\frac{\Xi^2\,\sigma_B^2}{P_B} \right)^{\frac{k_A+m_A}{2}}
    \nonumber \\ & \times x^{\frac{k_A+m_A}{2}-1}\, \G_{0,2}^{2,0}\left(\frac{\Xi^2\,\sigma_B^2}{2\,P_B} x \left| \frac{k_{A}-m_{A}}{2}, \frac{m_A-k_A}{2}\right.\right). 
    \label{Eq:f_gamma_B}
\end{align}
\begin{IEEEproof}
    For brevity, the proof of Lemma 1 is given in Appendix A. 
\end{IEEEproof}

\textbf{Lemma 2:} The CDF and PDF of $\gamma_E$ can be respectively obtained~as
\begin{align}
F_{\gamma_{E}}(x) = \sum_{n=1}^{N_{E}}  a_{n}^{E}\, c_{E}^{-b_{n}^{E}}\,\gamma\left(b_{n}^{E}, \frac{\sigma_E^{2}}{P_E} x\right)
\label{Eq:F_gamma_E}
\end{align}
and 
\begin{align}
    f_{\gamma_{E}}\left(x\right) = \sum_{n=1}^{N_{E}}  a_{n}^{E}\,
    \left(\frac{\sigma_E^{2}}{P_E c_{E}}\right)^{b_{n}^{E}}\, x^{b_{n}^{E}-1}\,\exp\left(-\frac{\sigma_E^{2}}{P_E} x \right).
    \label{Eq:f_gamma_E}
\end{align}
\begin{IEEEproof}
    For brevity, the proof of Lemma 2 is given in Appendix B. 
\end{IEEEproof}

\subsection{With full CSI-knowledge}\label{SS:Full_CSI}

In this section, we consider the ideal scenario in which Alice has full CSI-knowledge, i.e., both the instantaneous legitimate and eavesdropping channel coefficients are known by Alice. In this case, a transmission occurs only when the Alice-RIS-Bob channel is better than the Alice-Eve channel. As a consequence, the instantaneous secrecy rate is defined~as
\begin{align}
    C_s = \max\left(C_B-C_E, 0\right),
    \label{Eq:C_s}
\end{align}
where $C_B$ stands for the Alice-RIS-Bob link capacity, which can be obtained~as
\begin{align}
    C_B = \log_2\left(1+\gamma_{B}\right).
    \label{Eq:C_B}
\end{align} 
and $C_E$ represents the Alice-Eve link capacity that can be expressed~as
\begin{align}
    C_E = \log_2\left(1+\gamma_{E}\right).
    \label{Eq:C_E}
\end{align} 

\subsubsection{Probability of zero-secrecy capacity}\label{SS:PIL}
To assess the frequency of zero-secrecy capacity, we define the probability of zero secrecy capacity as
\begin{align}
    P_l = \Pr\left(C_E \geq C_B\right).
    \label{Eq:P_l_def}
\end{align}
The following theorem returns the probability of zero-secrecy capacity. 

\textbf{Theorem 1:} The probability of zero-secrecy capacity can be evaluated~as
\begin{align}
    P_l = \frac{1}{\Gamma\left(k_A\right)\Gamma\left(m_A\right)} &\sum_{n=1}^{N_{E}}  a_{n}^{E}\,
     c_{E}^{-b_{n}^{E}}\nonumber \\ & \hspace{-0.5cm} \times
    \G_{2,3}^{2,2}\left(\frac{\Xi^{2}\,\sigma_B^2\,P_{E}}{\sigma_{E}^{2}\,P_B}\left|\begin{array}{c} 1, 1-b_n^{E}\\ k_A, m_A, 0\end{array}\right.\right).
    \label{Eq:P_l}
\end{align}
\begin{IEEEproof}
    For brevity, the proof of Theorem 1 is given in Appendix C. 
\end{IEEEproof}

From~\eqref{Eq:P_l}, by accounting for the fact that $ \G_{2,3}^{2,2}\left(x\left|\begin{array}{c} 1, 1-b_n^{E}\\ k_A, m_A, 0\end{array}\right.\right)$ is an increasing function of $x$, we observe that for given $\Xi$ and $P_{B}/\sigma_{B}^2$, as $P_{E}/\sigma_E^2$ increases, the probability of information leakage increases. Likewise, for given $\Xi$ and $P_{E}/\sigma_{E}^2$, as $P_{B}/\sigma_B^2$ increases, the probability of information leakage decreases, which indicates that the PLS performance improve. Finally, for given $P_{B}/\sigma_{B}^2$ and $P_{E}/\sigma_{E}^2$, as $\Xi^2$ increases, the probability of information leakage also increases. Notice that $\Xi$ is a decreasing function of the number of MAs of the RIS. This indicates that as the number of MAs increases, the probability of information leakage decreases. 

\subsubsection{Average secrecy rate}\label{SS:AADR}



The following theorem returns the average secrecy rate of the wireless system for the case Alice have perfect and full CSI knowledge. 

\textbf{Theorem 2}:
For the case Alice have perfect and full CSI knowledge, the average achievable secrecy rate of the RIS-empowered wireless system can be obtained~as in~\eqref{Eq:av_Cs_final}, given at the top of the next page. 
\begin{figure*}
    \begin{align}
        \tilde{C}_s &=\frac{1}{\ln(2)}
    \frac{1}{\Gamma\left(k_A\right)\Gamma\left(m_A\right)} 
    \left(\frac{\Xi^2\, P_E\,\sigma_B^2}{P_B\,\sigma_E^2} \right)^{\frac{k_A+m_A}{2}} 
    \nonumber \\ & \times
    \sum_{n=1}^{N_{E}} a_{n}^{E}\, c_{E}^{-b_{n}^{E}}\, 
    \G_{1,2:2,2:0,0}^{1,1:1,2:2,0}\left( 
        \begin{array}{c}  
            1-\frac{k_A+m_A}{2}-b_n^{E}, 1-\frac{k_A+m_A}{2} \\ -\frac{k_A+m_A}{2}
        \end{array}
        \left| 
         \begin{array}{c}  
            1, 1 \\ 1, 0
        \end{array}
        \right.
        \left| 
        \begin{array}{c}  
             - \\
             \frac{k_A-m_A}{2}, \frac{m_A-k_A}{2}
        \end{array}
        \right|
        \frac{P_E}{\sigma_E^2}, \frac{\Xi^2\,\sigma_{B}^{2}\,P_E}{2\,P_B\,\sigma_{E}^{2}}
    \right)
    \nonumber \\ & - \frac{1}{\ln(2)}\frac{1}{\Gamma\left(k_A\right)\Gamma\left(m_A\right)}\sum_{n=1}^{N_{E}}  a_{n}^{E}\,
    \left(\frac{\sigma_E^{2}}{P_E c_{E}}\right)^{b_{n}^{E}}\, \G_{1,0:2,2:1,3}^{0,1:1,2:2,1}\left( 1-b_n^{E} \left|\begin{array}{c} 1, 1 \\ 1, 0 \end{array}\right|\left.\begin{array}{c} 1\\ k_A, m_A, 0 \end{array}\right| \frac{P_E}{\sigma_E^2}, \frac{\Xi^2 \sigma_B^2 P_E}{\sigma_E^2\,P_B} \right)
    \label{Eq:av_Cs_final}
    \end{align}
    \hrulefill
\end{figure*}
\begin{IEEEproof}
For brevity, the proof of Theorem 2 is provided in Appendix D. 
\end{IEEEproof}

\subsection{With partial CSI knowledge}\label{SS:PCSI}

In this section, we assume that the CSI of the Alice-Eve channel is also unknown to Alice. In more detail, it is assumed that there is an area of trust that includes Alice, the RIS, and Bob, while Eve is outside this zone. Notice that, according to~\cite{7543509,Trevlakis2023,9625032}, this is expected to be a realistic scenario for sixth generation wireless systems. In this case, Alice continuously sends information messages to Bob independently of the legitimate and eavesdropper channel conditions. Thus, in this case, the instantaneous secrecy capacity is defined~as 
\begin{align}
	R_{s} = C_{B}-C_{E}.
	\label{Eq:Rs_instance}
\end{align}
From~\eqref{Eq:Rs_instance}, it is evident that $R_s$ can instantaneously take negative values, in this case an information leakage occurs.

\subsubsection{Probability of information leakage}
The probability of information leakage is defined as
\begin{align}
    P_i = \Pr\left(R_s\leq 0\right),
\end{align}
which, by applying~\eqref{Eq:Rs_instance}, can be written~as
\begin{align}
    P_i = \Pr\left(C_B\leq C_E\right).
    \label{Eq:Pi}
\end{align}
By comparing~\eqref{Eq:Pi} with~\eqref{Eq:P_l_def}, it becomes evident that the probability of information leakage for the case of average CSI knowledge is the same as the probability of zero-secrecy capacity for the case of instantaneous CSI knowledge; thus, it can be obtained as in~\eqref{Eq:P_l}.

\subsubsection{Average  secrecy  rate}
The following theorem returns a closed-form expression for the average secrecy rate for the case Alice has no knowledge of the instantaneous characteristics of the Alice-Eve channel. 

\textbf{Theorem 3:} For the case Alice has no knowledge of the instantaneous characteristics of the Alice-Eve channel, the average secrecy rate can be obtained~as in~\eqref{Eq:R_s_final}, given at the top of the next page. 
	\begin{figure*}
		\begin{align}
			\tilde{R}_s &=\frac{1}{\ln(2)}
   \left(\frac{P_B}{\sigma_B^2}\right)^{-\frac{k_A+m_A}{2}}  \frac{ \Xi^{k_A + m_A}}{\Gamma\left(k_A\right)\Gamma\left(m_A\right)}
			 \mathrm{G}_{2,4}^{4,1}\left(\frac{\Xi^2\,\sigma_B^2}{P_B}\left|\begin{array}{c} -\frac{k_A+m_A}{2}, 1-\frac{k_A+m_A}{2}\\ \frac{k_A-m_A}{2}, - \frac{k_A-m_A}{2}, -\frac{k_A+m_A}{2}, -\frac{k_A+m_A}{2}\end{array}\right.\right)
			\nonumber \\ & 
			- \sum_{n=1}^{N_{E}} \frac{ a_{n}^{E}\, c_E^{-b_n^{E}}}{\ln(2)} 
			\mathrm{G}_{4,3}^{1,4}\left(\frac{P_E}{c_E\, \sigma_E^{2}}\left|\begin{array}{c} 0, 0, 1-b_n^{E}, -b_n^{E}\\ 0, -b_n^{E}, -1 \end{array}\right.\right)
			\label{Eq:R_s_final}
		\end{align} 
		\hrulefill
	\end{figure*}
 \begin{IEEEproof}
For brevity, the proof of Theorem 3 is provided in Appendix E. 
\end{IEEEproof}

\section{Numerical results and discussions}\label{S:NR}

The objective of this section is to verify the theoretical framework through Monte Carlo simulations and to present numerical results that highlight the PLS capabilities of RIS-empowered wireless systems. The following scenario is considered.  We assume that the Alice-RIS and RIS-Bob channel coefficients follow independent and identical Rice distributions with $K_r$ parameter equal to $5\,\mathrm{dB}$, while the Alice-Eve channel coefficient is modeled as a Nakagami-$m$ RV with $m=3$. For an accurate approximation of the Rice distribution, we select $N_A=N_R=N_r=20$, 
\begin{align}
	a_{n}^{\mathcal{X}} = \frac{\delta\left(K_r, n\right)}{\sum_{k_1=1}^{N_r} \delta\left(K_r, k_1\right) \Gamma\left(b_{k_1}^{\mathcal{X}}\right) c_{\mathcal{X}}^{-b_{k_1}^{\mathcal{X}}}},
	\label{Eq:a_k_rice}
\end{align}    
where
\begin{align}
	\delta\left(K_r, n\right) = \frac{K_r^{n-1}\left(1+K_r\right)^n}{\exp\left(K_1\right) \left( (n-1)!\right)^2},
\end{align}
while 
\begin{align}c_{\mathcal{X}}=1+K_r\end{align} 
and 
\begin{align}
b_{n}^{\mathcal{X}}=n.
\end{align} 
Note that $\mathcal{X}\in\{A, R\}$ and
\begin{align}
	N_r=\left\{\begin{array}{c}M, \, i=1 \\ K, \, i=2 \end{array}\right..
\end{align}
The MG distribution can be simplified to Nakagami-$m$ by setting $N_{E}=1$, 
\begin{align}
a_n^{E}=\frac{m^{m}}{\Gamma(m)}
\end{align}
and 
\begin{align}
b_n^{E}=c_3=m, 
\end{align}
with $m$ being the shape parameter. 
In what follows, lines are used to denote analytical results while markers simulations. 

\begin{figure}
	\centering
	\scalebox{1.00}{\begin{tikzpicture}
\begin{axis}[
	xlabel={$\frac{\rho_B}{\rho_{E}}\,\mathrm{(dB)}$},
	ylabel={$P_l$},
    ymode = log,
	legend pos=north east,
	xmin = -14,
	xmax = 12,
	ymin = 10^-7,
	ymax= 10^-3,
	xtick = {-14, -12, -10, -8, -6, -4, -2, 0, 2, 4, 6, 8, 10},
	ymajorgrids=true,
	xmajorgrids=true,
	grid style=dashed,
	]
	
	\addplot[
	color=black]
	coordinates {
        (-10, 1.0)
        (-9, 1.0)
        (-8, 0.379986)
        (-7, 0.115206)
        (-6, 0.0712851)
        (-5, 0.0421649)
        (-4, 0.0235687)
        (-3, 0.0124664)
        (-2, 0.00625434)
        (-1, 0.00298492)
        (0, 0.00135979)
        (1, 0.000593484)
        (2, 0.000249132)
        (3, 0.000100982)
        (4, 0.0000396767)
        (5, 0.0000151682)
        (6, 5.66203*10^-6)
        (7, 2.07052*10^-6)
        (8, 7.43971*10^-7)
        (9, 2.63372*10^-7)
        (10, 9.20776*10^-8)
    };
    \addlegendentry{\small{Analytical}}

    \addplot[
	    color=black,
        only marks,
	    mark = square,]
	coordinates {
        (-10, 1.0)
        (-9, 1.0)
        (-8, 0.379986)
        (-7, 0.115206)
        (-6, 0.0712851)
        (-5, 0.0421649)
        (-4, 0.0235687)
        (-3, 0.0124664)
        (-2, 0.00625434)
        (-1, 0.00298492)
        (0, 0.00135979)
        (1, 0.000593484)
        (2, 0.000249132)
        (3, 0.000100982)
        (4, 0.0000396767)
        (5, 0.0000151682)
        (6, 5.66203*10^-6)
        (7, 2.07052*10^-6)
        (8, 7.43971*10^-7)
        (9, 2.63372*10^-7)
        (10, 9.20776*10^-8)
    };
    \addlegendentry{\small{Sim. (M=4)}}


    \addplot[
	   color=black,
        only marks,
	   mark = *]
	coordinates {
        (-10.0, 1.00)
        (-9.15, 0.733255)
        (-9.1, 0.547616)
        (-9.0, 0.294035)
        (-8.9, 0.16999)
        (-8.8, 0.0877033)
        (-8.75, 0.0689474)
        (-8.7, 0.0544684)
        (-8.6, 0.0306695)
        (-8.5, 0.0182641)
       (-8.35, 0.00886667)
        (-8.0, 0.00254969)
        (-7.5, 0.00105426)
        (-7.0, 0.000572983)
        (-6.5, 0.000318264)
        (-6.0, 0.000173403)
        (-5.5, 0.0000923254)
        (-5.0, 0.0000480295)
        (-4.5, 0.0000244229)
        (-4.0, 0.0000121454)
        (-3.5, 5.91005*10^-6)
        (-3.0, 2.81574*10^-6)
        (-2.5, 1.31425*10^-6)
        (-2.0, 6.01341*10^-7)
        (-1.5, 2.69893*10^-7)
        (-1.0, 1.18897*10^-7)
        (-0.5, 5.14443*10^-8)
    };
    \addlegendentry{\small{Sim. (M=8)}}

    \addplot[
	   color=black,
        only marks,
	   mark = o]
	coordinates {
        (-16, 0.032963)
        (-14, 0.00948735)
        (-12, 0.00221849)
        (-10, 0.000429345)
        (-8.0, 0.0000702865)
        (-6.0, 9.95566*10^-6)
        (-4.0, 1.24685*10^-6)
        (-3.0, 4.23994*10^-7)
        (-2.0, 1.40826*10^-7)
        (-1.5, 8.05007*10^-8)
    };
    \addlegendentry{\small{Sim. (M=16)}}

    \addplot[
	   color=black,
       only marks,
	   mark = triangle]
	coordinates {
        (-15.0, 0.0000350409)
        (-14.0, 0.0000110539)
        (-13.5, 6.09203*10^-6)
        (-13.0, 3.31735*10^-6)
        (-12.0, 9.50565*10^-7)
        (-11.0, 2.60995*10^-7)
        (-10.0, 6.89015*10^-8)
    };
    \addlegendentry{\small{Sim. (M=32)}}

    \addplot[
	color=black]
	coordinates {
        (-10.0, 1.00)
        (-9.2, 1.00)
        (-9.15, 0.733255)
        (-9.1, 0.547616)
        (-9.0, 0.294035)
        (-8.9, 0.16999)
        (-8.8, 0.0877033)
        (-8.75, 0.0689474)
        (-8.7, 0.0544684)
        (-8.6, 0.0306695)
        (-8.5, 0.0182641)
        (-8.4, 0.011465)
        (-8.35, 0.00886667)
        (-8.3, 0.00746705)
        (-8.2, 0.00494726)
        (-8.1, 0.00351547)
        (-8.0, 0.00254969)
        (-7.9, 0.00206042)
        (-7.8, 0.00165956)
        (-7.7, 0.00139796)
        (-7.6, 0.00120799)
        (-7.5, 0.00105426)
        (-7.4, 0.000924634)
        (-7.3, 0.00081765)
        (-7.2, 0.000725663)
        (-7.1, 0.000643957)
        (-7.0, 0.000572983)
        (-6.9, 0.000510029)
        (-6.8, 0.000453688)
        (-6.7, 0.000403382)
        (-6.6, 0.000358441)
        (-6.5, 0.000318264)
        (-6.4, 0.000282347)
        (-6.3, 0.000250302)
        (-6.2, 0.000221679)
        (-6.1, 0.000196148)
        (-6.0, 0.000173403)
        (-5.9, 0.00015315)
        (-5.8, 0.000135139)
        (-5.7, 0.000119133)
        (-5.6, 0.000104925)
        (-5.5, 0.0000923254)
        (-5.4, 0.0000811635)
        (-5.3, 0.0000712848)
        (-5.2, 0.0000625507)
        (-5.1, 0.0000548365)
        (-5.0, 0.0000480295)
        (-4.9, 0.0000420292)
        (-4.8, 0.0000367452)
        (-4.7, 0.0000320964)
        (-4.6, 0.0000280105)
        (-4.5, 0.0000244229)
        (-4.4, 0.0000212758)
        (-4.3, 0.0000185178)
        (-4.2, 0.0000161032)
        (-4.1, 0.0000139911)
        (-4.0, 0.0000121454)
        (-3.9, 0.000010534)
        (-3.8, 9.12853*10^-6)
        (-3.7, 7.90378*10^-6)
        (-3.6, 6.83751*10^-6)
        (-3.5, 5.91005*10^-6)
        (-3.4, 5.10408*10^-6)
        (-3.3, 4.40432*10^-6)
        (-3.2, 3.79732*10^-6)
        (-3.1, 3.27125*10^-6)
        (-3.0, 2.81574*10^-6)
        (-2.5, 1.31425*10^-6)
        (-2.0, 6.01341*10^-7)
        (-1.5, 2.69893*10^-7)
        (-1.0, 1.18897*10^-7)
        (-0.5, 5.14443*10^-8)
    };

    \addplot[
	   color=black]
	coordinates {
        (-16, 0.032963)
        (-14, 0.00948735)
        (-12, 0.00221849)
        (-10, 0.000429345)
        (-8.0, 0.0000702865)
        (-6.0, 9.95566*10^-6)
        (-4.0, 1.24685*10^-6)
        (-3.0, 4.23994*10^-7)
        (-2.0, 1.40826*10^-7)
        (-1.5, 8.05007*10^-8)
    };

    \addplot[
	   color=black]
	coordinates {
        (-15.0, 0.0000350409)
        (-14.0, 0.0000110539)
        (-13.5, 6.09203*10^-6)
        (-13.0, 3.31735*10^-6)
        (-12.0, 9.50565*10^-7)
        (-11.0, 2.60995*10^-7)
        (-10.0, 6.89015*10^-8)
    };


\end{axis}
\end{tikzpicture}}
	\vspace{-0.25cm}
	\caption{$P_l$ vs $\frac{\rho_B}{\rho_E}$ for different values of $M$.}
	\label{Fig:Prob_leak}
\end{figure}
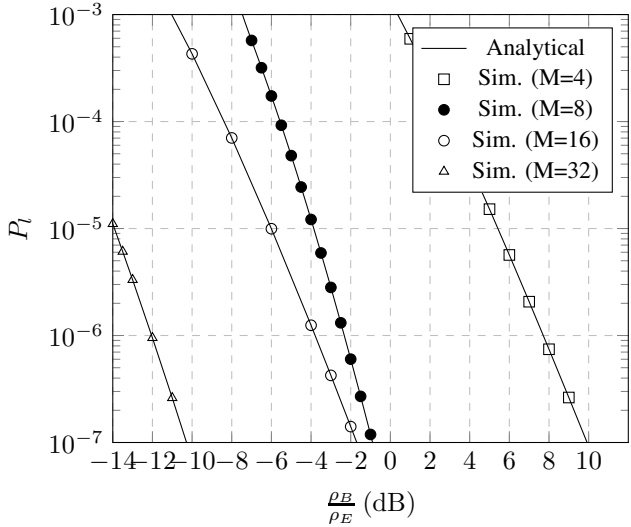

Figure~\ref{Fig:Prob_leak} depicts the probability of information leakage or the probability of zero-secrecy data rate as a function of $\frac{\rho_B}{\rho_E}$ for different values of $M$. As expected, for a given $M$, as $\frac{\rho_B}{\rho_E}$ increases, the probability of information leakage decreases. For example for $M=4$, the probability of information leakage decreases from $1.3\times 10^{-3}$ to $2.49\times 10^{-4}$, as $\frac{\rho_B}{\rho_E}$ increases from $0$ to $2\,\mathrm{dB}$. Likewise, for a fixed $\frac{\rho_B}{\rho_E}$, as $M$ increases the probability of information leakage or the probability of zero-secrecy data rate decreases.  



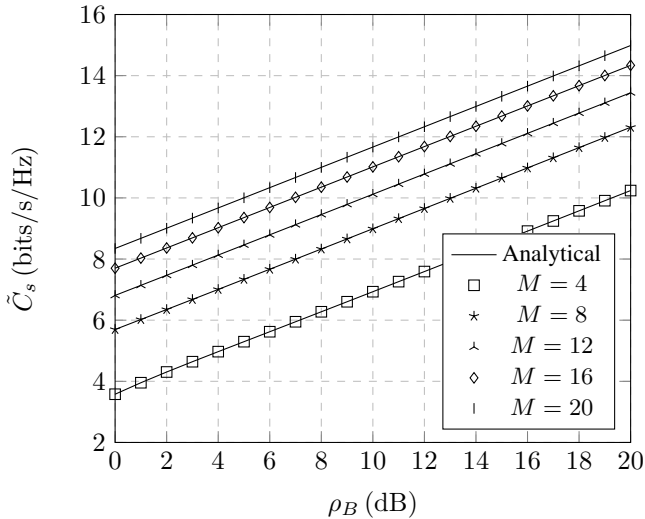
\begin{figure}
	\centering
	\scalebox{1.00}{\begin{tikzpicture}
\begin{axis}[
	xlabel={$\rho_B\,\mathrm{(dB)}$},
	ylabel={$\tilde{C}_s\,\mathrm{(bits/s/Hz)}$},
	legend pos=south east,
	xmin = 0,
	xmax =20,
	ymin = 2,
	ymax= 16,
	xtick = {0, 2, 4, 6, 8, 10, 12, 14, 16, 18, 20},
	ytick = {2, 4, 6, 8, 10, 12, 14, 16},
	ymajorgrids=true,
	xmajorgrids=true,
	grid style=dashed,
	]

\addplot[
    color=black]
coordinates {
	(0, 3.57771)
    (1, 3.95372)
    (2, 4.30479)
    (3, 4.64127)
    (4, 4.97041)
    (5, 5.29659)
    (6, 5.62215)
    (7, 5.94817)
    (8, 6.27505)
    (9, 6.60283)
    (10, 6.93143)
    (11, 7.26074)
    (12, 7.59062)
    (13, 7.92097)
    (14, 8.2517)
    (15, 8.58272)
    (16, 8.91399)
    (17, 9.24544)
    (18, 9.57705)
    (19, 9.90877)
    (20, 10.2406)
};
\addlegendentry{\small{Analytical}}

\addplot[
	    color=black,
        only marks,
	    mark = square,]
	coordinates {
        (0, 3.57771)
        (1, 3.95372)
        (2, 4.30479)
        (3, 4.64127)
        (4, 4.97041)
        (5, 5.29659)
        (6, 5.62215)
        (7, 5.94817)
        (8, 6.27505)
        (9, 6.60283)
        (10, 6.93143)
        (11, 7.26074)
        (12, 7.59062)
        (13, 7.92097)
        (14, 8.2517)
        (15, 8.58272)
        (16, 8.91399)
        (17, 9.24544)
        (18, 9.57705)
        (19, 9.90877)
        (20, 10.2406)
    };
\addlegendentry{\small{$M=4$}}


\addplot[
	    color=black,
        only marks,
	    mark = star,]
	coordinates {
        (0, 5.69573)
    (1, 6.02178)
    (2, 6.34907)
    (3, 6.67736)
    (4, 7.00644)
    (5, 7.33616)
    (6, 7.66638)
    (7, 7.997)
    (8, 8.32795)
    (9, 8.65915)
    (10, 8.99055)
    (11, 9.32212)
    (12, 9.65381)
    (13, 9.98561)
    (14, 10.3175)
    (15, 10.6494)
    (16, 10.9814)
    (17, 11.3135)
    (18, 11.6455)
    (19, 11.9776)
    (20, 12.3097)
    };
\addlegendentry{\small{$M=8$}}


\addplot[
	    color=black,
        only marks,
	    mark = Mercedes star]
	coordinates {
        (0, 6.81048)
        (1, 7.13972)
        (2, 7.46956)
        (3, 7.79987)
        (4, 8.13057)
        (5, 8.46158)
        (6, 8.79283)
        (7, 9.12427)
        (8, 9.45587)
        (9, 9.78759) 
        (10, 10.1194)
        (11, 10.4513)
        (12, 10.7833)
        (13, 11.1153)
        (14, 11.4473)
        (15, 11.7794)
        (16, 12.1115)
        (17, 12.4436)
        (18, 12.7757)
        (19, 13.1079)
        (20, 13.44)
    };
\addlegendentry{\small{$M=12$}}


\addplot[
	    color=black,
        only marks,
	    mark = diamond]
	coordinates {
        (0, 7.70004)
        (1, 8.03076)
        (2, 8.36178)
        (3, 8.69305)
        (4, 9.0245)
        (5, 9.3561) 
        (6, 9.68783)
        (7, 10.0196)
        (8, 10.3515)
        (9, 10.6835)
        (10, 11.0155)
        (11, 11.3476)
        (12, 11.6796)
        (13, 12.0117)
        (14, 12.3439)
        (15, 12.676)
        (16, 13.0081)
        (17, 13.3403)
        (18, 13.6724)
        (19, 14.0046)
        (20, 14.3368)
    };
\addlegendentry{\small{$M=16$}}


\addplot[
	    color=black,
        only marks,
	    mark = |]
	coordinates {
        (0, 8.34612)
        (1, 8.67738)
        (2, 9.00883)
        (3, 9.34044)
        (4, 9.67216)
        (5, 10.004)
        (6, 10.3359)
        (7, 10.6678)
        (8, 10.9998)
        (9, 11.3319)
        (10, 11.664)
        (11, 11.9961)
        (12, 12.3282)
        (13, 12.6603)
        (14, 12.9925)
        (15, 13.3246)
        (16, 13.6568)
        (17, 13.9889)
        (18, 14.3211)
        (19, 14.6533)
        (20, 14.9855)
    };
\addlegendentry{\small{$M=20$}}

\addplot[
    color=black]
coordinates {
	(0, 5.69573)
    (1, 6.02178)
    (2, 6.34907)
    (3, 6.67736)
    (4, 7.00644)
    (5, 7.33616)
    (6, 7.66638)
    (7, 7.997)
    (8, 8.32795)
    (9, 8.65915)
    (10, 8.99055)
    (11, 9.32212)
    (12, 9.65381)
    (13, 9.98561)
    (14, 10.3175)
    (15, 10.6494)
    (16, 10.9814)
    (17, 11.3135)
    (18, 11.6455)
    (19, 11.9776)
    (20, 12.3097)
};


\addplot[
	    color=black]
	coordinates {
        (0, 6.81048)
        (1, 7.13972)
        (2, 7.46956)
        (3, 7.79987)
        (4, 8.13057)
        (5, 8.46158)
        (6, 8.79283)
        (7, 9.12427)
        (8, 9.45587)
        (9, 9.78759) 
        (10, 10.1194)
        (11, 10.4513)
        (12, 10.7833)
        (13, 11.1153)
        (14, 11.4473)
        (15, 11.7794)
        (16, 12.1115)
        (17, 12.4436)
        (18, 12.7757)
        (19, 13.1079)
        (20, 13.44)
    };


    \addplot[
	    color=black]
	coordinates {
        (0, 7.70004)
        (1, 8.03076)
        (2, 8.36178)
        (3, 8.69305)
        (4, 9.0245)
        (5, 9.3561) 
        (6, 9.68783)
        (7, 10.0196)
        (8, 10.3515)
        (9, 10.6835)
        (10, 11.0155)
        (11, 11.3476)
        (12, 11.6796)
        (13, 12.0117)
        (14, 12.3439)
        (15, 12.676)
        (16, 13.0081)
        (17, 13.3403)
        (18, 13.6724)
        (19, 14.0046)
        (20, 14.3368)
    };

    \addplot[
	    color=black]
	coordinates {
        (0, 8.34612)
        (1, 8.67738)
        (2, 9.00883)
        (3, 9.34044)
        (4, 9.67216)
        (5, 10.004)
        (6, 10.3359)
        (7, 10.6678)
        (8, 10.9998)
        (9, 11.3319)
        (10, 11.664)
        (11, 11.9961)
        (12, 12.3282)
        (13, 12.6603)
        (14, 12.9925)
        (15, 13.3246)
        (16, 13.6568)
        (17, 13.9889)
        (18, 14.3211)
        (19, 14.6533)
        (20, 14.9855)
    };

\end{axis}

\end{tikzpicture}}
	\vspace{-0.25cm}
	\caption{$\tilde{C}_s$ vs $\rho_B$ for different values of $M$, assuming $\rho_E=0\,\rm{dB}$ and full CSI knowledge.}
	\label{Fig:C_s}
\end{figure}

Figure~\ref{Fig:C_s} illustrates the average secrecy rate as a function of $\rho_B$ for different values of $M$, assuming $\rho_E=0\,\rm{dB}$ and full CSI knowledge. In this figure, continuous lines denote theoretical results, while the markers stands for simulations. It becomes evident that the theoretical framework and the simulations match; thus, the theoretical framework is verified. Additional, we observe that for a given $M$, as $\rho_B$ increases, the average secrecy rate also increases. For instance, for $M=16$, the  average secrecy rate increases from $11.0$ to $11.7\,\rm{bits/s/Hz}$, as $\rho_B$ increases from $10$ to $12\,\rm{dB}$. Moreover, for a fixed $\rho_B$, we observe that as $M$ increases, the diversity order and gain of the legitimate link increase; hence, the average secrecy rate increases. For example, for $\rho_B=10\,\rm{dB}$, as $M$ increases from $4$ to $8$, the average secrecy rate increases for approximately $0.85\,\rm{bits/s/Hz}$, whereas, for the same $\rho_B$, the average secrecy rate increases for about $2.03\,\rm{bits/s/Hz}$, as $M$ increases from $8$ to $16$.



\begin{figure}
	\centering
	\scalebox{1.00}{\input{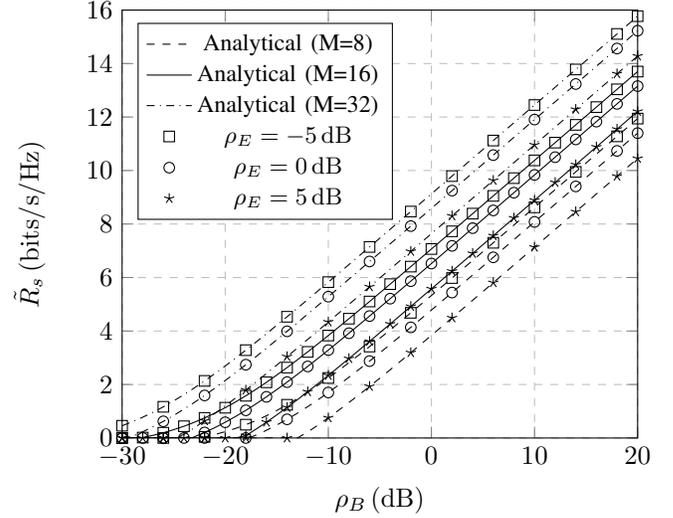}}
	\vspace{-0.25cm}
	\caption{$\tilde{R}_s$ vs $\rho_B$, for different values of $\rho_E$ and $M$.}
	\label{Fig:PlotFig}
\end{figure} 

In Fig.~\ref{Fig:PlotFig}, the average secrecy rate is illustrated as a function of $\rho_B$, for different values of $\rho_E$ and $M$, for the case of average CSI knowledge. We observe that for given $\rho_{E}$ and $M$, as $\rho_B$ increases, the legitimate channel improves; as a result, the average secrecy rate increases. For example, for $\rho_E=0\,\mathrm{dB}$  and $M=32$, the average secrecy rate increases from $5.28$ to $8.59$, as $\rho_B$ increases from $-10$ to $0\,\mathrm{dB}$. Likewise, for fixed $\rho_{B}$ and $M$, as $\rho_{E}$ increases, the average secrecy rate decreases. For instance, for $\rho_B=0\,\mathrm{dB}$ and $N=32$, the average secrecy rate changes from $8.59$ to $7.64\,\mathrm{bits/s/Hz}$, as $\rho_{E}$ increases from $0$ to $5\,\mathrm{dB}$. Finally, for given $\rho_B$ and  $\rho_{E}$, as $M$ increases, the diversity order and gain increase; hence, the average secrecy rate increases.

\begin{figure}
	\centering
	\scalebox{1.00}{\begin{tikzpicture}
	\begin{axis}[
		legend columns=1,
		legend entries={$\rho_{E}=-5\,\mathrm{dB}$, $\rho_{E}=0\,\mathrm{dB}$, 
        },
		ybar legend,
        bar width = 4pt,
		ymajorgrids=true,
		legend pos = north west,
		xlabel = {$M$},
		xtick={4, 8, 12, 16, 20, 24, 28},
		ylabel ={Average secrecy data rate $\mathrm{(bits/s/Hz)}$},
		]
		
		\addplot[ybar, pattern=north east lines] coordinates {
			(4.0, 3.34272) 
			(6.0, 4.507684643969879)
			(8.0, 5.327193755602597)
			(10.0, 5.780307297325157)
			(12.0, 6.257356290460929)
			(14.0, 6.68391028212764)
			(16.0, 7.066345365356715)
			(18.0, 7.424168116566778)
			(20.0, 7.732992038083899)
			(22.0, 8.01434235305475)
			(24.0, 8.272335859659929)
			(26.0, 8.51032281490735)
			(28.0, 8.731034537612176)
			(30.0, 8.936710677446557)
		};
	\addplot[ybar, pattern=horizontal lines] coordinates {
		(4.0, 2.80787) 
		(6.0, 3.9639555649364207)
		(8.0, 4.78346)
		(10.0, 5.236578218291699)
		(12.0, 5.713627211427471)
		(14.0, 6.1401812030941825)
		(16.0, 6.522616286323257)
		(18.0, 6.88043903753332)
		(20.0, 7.189262959050441)
		(22.0, 7.470613274021294)
		(24.0, 7.728606780626472)
		(26.0, 7.966593735873894)
		(28.0, 8.18730545857872)
		(30.0, 8.392981598413101)
	};

 \addplot[ybar, blue, fill=white, pattern=north east lines,  pattern color=.] coordinates {
			(5.0, 5.29659)
            (7.0, 6.49328)
            (9.0, 7.33616)
            (11.0, 7.98793)
            (13.0, 8.46158)
            (15.0, 8.96794)
            (17.0, 9.3561)
            (19.0, 9.69819)
            (21.0, 10.004)
            (23.0, 10.2804)
            (25.0, 10.5327)
            (27.0, 10.7647)
            (29.0, 10.9794)
            (31.0, 11.1)
		};

    \addplot[ybar, blue, fill=white, pattern=horizontal lines,  pattern color=.] coordinates {
			(5.0, 3.57771) 
			(7.0, 4.86794)
            (9.0, 5.69573)
            (11.0, 6.33987)
            (13.0, 6.81048)
            (15.0, 7.31342)
            (17.0, 7.70004)
            (19.0, 8.04108)
            (21.0, 8.34612)
            (23.0, 8.62204)
            (25.0, 8.87391)
            (27.0, 9.10557)
            (29.0, 9.32002)
            (31.0, 9.53)
		};

	\end{axis}

\end{tikzpicture}}
	\vspace{-0.25cm}
	\caption{Average secrecy rate vs $M$ for different values of $\rho_E$, assuming $\rho_B=0\,\mathrm{dB}$, full (blue) and partial (black) CSI knowledge.}
	\label{Fig:RS_to_M}
\end{figure}
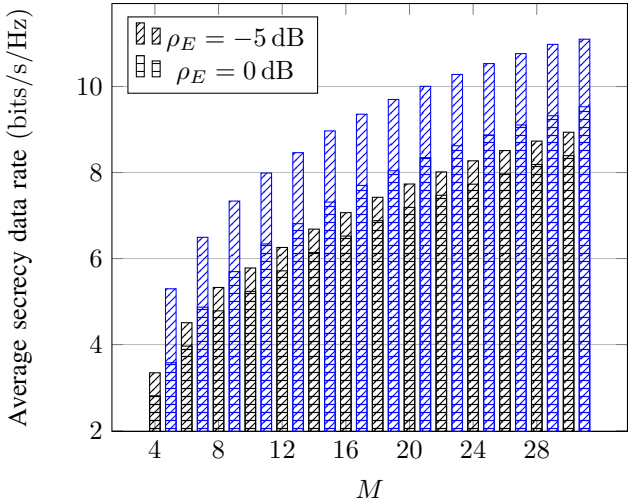 

In Fig.~\ref{Fig:RS_to_M}, the average secrecy rate is presented as a function of the number of RIS's MAs, $M$, for different values of $\rho_E=\frac{P_{E}}{\sigma_{E}^{2}}$, assuming that $\rho_B=\frac{P_B}{\sigma_B^2}=0$. With black color, the scenario in which average CSI knowledge is depicted, while black color is used for the case of full CSI knowledge. From this figure, we observe that, for a given $\rho_{E}$ and a specific scenario, as $M$ increases the average secrecy rate also increases. For instance, for the case of average CSI knowledge, for $\rho_{E}=0\text{}\mathrm{dB}$, the average secrecy rate increases from $2.81$ to $3.96\,\mathrm{dB}$ as $M$ increases from $4$ to $6$. In other words, an increase of approximately $1.15\,\mathrm{dB}$ is observed. Moreover, for the same average secrecy rate, as $M$ increases from $6$ to $8$, a $0.82\,\mathrm{dB}$ increase is observed. This indicates that as $M$ increases, the gain in terms of average secrecy rate decreases. Likewise, from this figure, it becomes evident that, for a given $M$, as $\rho_{E}$ increases, the average secrecy rate decreases. Finally, for the same $M$ and $\rho_E$, we observe that the scenario with full CSI knowledge outperforms the one with partial CSI knowledge in terms of average secrecy rate.

\section{Conclusions}\label{S:Con}

In this paper, we examined the PLS capabilities of RIS-empowered wireless systems with full and partial CSI knowledge. In this direction, we presented closed-form expressions for the probability of zero-secrecy capacity and the probability of information leakage as well as the average achievable secrecy rate. Our results revealed that there is a relationship between the size of the RIS and the PLS capabilities. In more detail, as the RIS size increases, the PLS capabilities of the system improves. Finally, it was highlighted that assuming perfect CSI knowledge may lead to importantly overestimating the performance of the system. 

\section*{Appendices}

\section*{Appendix A}
\section*{Proof of Lemma 1}

The CDF of $\gamma_B$ is defined~as
\begin{align}
    F_{\gamma_{B}}\left(x\right) = \Pr\left(\gamma_B\leq x\right).
    \label{Eq:F_gamma_B_def}
\end{align}
By applying~\eqref{Eq:gamma_B},~\eqref{Eq:F_gamma_B_def} can be rewritten~as
\begin{align}
    F_{\gamma_{B}}\left(x\right) = \Pr\left( \frac{A_r^{2}\,P_B}{\sigma_B^2} \leq x \right)
\end{align}
or equivalently
\begin{align}
    F_{\gamma_{B}}\left(x\right) = \Pr\left( A_r \leq \sqrt{\frac{x\,\sigma_B^2}{P_B}}\right),
\end{align}
which can be rewritten~as
\begin{align}
    F_{\gamma_{B}}\left(x\right) = F_{A_r}\left( \sqrt{\frac{x\,\sigma_B^2}{P_B}}\right).
    \label{Eq:F_gamma_B_step_3}
\end{align}
By applying~\eqref{Eq:F_A} in~\eqref{Eq:F_gamma_B_step_3}, we obtain~\eqref{Eq:F_gamma_B}.

To evaluate the PDF of $\gamma_B$, we calculate the derivative of~\eqref{Eq:F_gamma_B}~as
\begin{align}
    f_{\gamma_{B}}\left(x\right) = \frac{\mathrm{d}F_{\gamma_{B}}\left(x\right)}{\mathrm{dx}},
\end{align}
or equivalently 
\begin{align}
f_{\gamma_{B}}\left(x\right) &= \frac{\mathrm{d}}{\mathrm{dx}} \left(\frac{1}{\Gamma\left(k_A\right)\Gamma\left(m_A\right)} \right. \nonumber \\ &\times \left. \mathrm{G}_{1, 3}^{2, 1}\left(\Xi^2 \frac{x\,\sigma_B^2}{P_B}\left|\begin{array}{c} 1\\ k_A, m_A, 0 \end{array}\right.\right) \right), 
\end{align}
or
\begin{align}
f_{\gamma_{B}}\left(x\right) &= \frac{1}{\Gamma\left(k_A\right)\Gamma\left(m_A\right)} 
\nonumber \\ & \times\frac{\mathrm{d}}{\mathrm{dx}} \left( \mathrm{G}_{1, 3}^{2, 1}\left(\Xi^2 \frac{x\,\sigma_B^2}{P_B}\left|\begin{array}{c} 1\\ k_A, m_A, 0 \end{array}\right.\right) \right), 
\label{Eq:f_gamma_B_step_3}
\end{align}
which can be rewritten as
\begin{align}
    f_{\gamma_{B}}\left(x\right) &= \frac{2}{\Gamma\left(k_A\right)\Gamma\left(m_A\right)} \left(\frac{\Xi^2\,\sigma_B^2}{P_B} \right)^{\frac{k_A+m_A}{2}}
    \nonumber \\ & \times x^{\frac{k_A+m_A}{2}-1}\, \K_{m_A-k_A}\left(2\sqrt{\frac{\Xi^2\,\sigma_B^2}{P_B} x} \right).
\end{align}
Finally, by employing~\cite[Eq.(9.34/4)]{B:Gra_Ryz_Book}, we obtain~\eqref{Eq:f_gamma_B}. This concludes the~proof.

\section*{Appendix B}
\section*{Proof of Lemma 2}
The CDF of $\gamma_{E}$ is defined~as
\begin{align}
    F_{\gamma_{E}}\left(x\right) = \Pr\left(\gamma_{E}\leq x\right).
    \label{Eq:F_gamma_E_def}
\end{align}
By applying~\eqref{Eq:gamma_E},~\eqref{Eq:F_gamma_E_def} can be written~as
\begin{align}
    F_{\gamma_{E}}\left(x\right) = \Pr\left(\frac{\left|h_{E}\right|^{2}\,P_E}{\sigma_{E}^{2}}\leq x \right),
\end{align}
or equivalently
\begin{align}
    F_{\gamma_{E}}\left(x\right) = \Pr\left(\left|h_{E}\right| \leq \sqrt{\frac{\sigma_E^{2}}{P_E} x} \right),
\end{align}
or
\begin{align}
    F_{\gamma_{E}}\left(x\right) = F_{h_E}\left(\sqrt{\frac{\sigma_E^{2}}{P_E} x} \right). 
    \label{Eq:F_gamma_E_step_3}
\end{align}
With the aid of~\eqref{Eq:F_h_E},~\eqref{Eq:F_gamma_E_step_3} can be written as in~\eqref{Eq:F_gamma_E}. 

To find the PDF of $\gamma_E$, we calculate the derivative of~\eqref{Eq:F_gamma_E}~as
\begin{align}
    f_{\gamma_{E}}\left(x\right) = \frac{\mathrm{d}F_{\gamma_{E}}\left(x\right)}{\mathrm{dx}},
\end{align}
which, with the aid of~\eqref{Eq:F_gamma_E}, can be rewritten~as
\begin{align}
    f_{\gamma_{E}}\left(x\right) = \frac{\mathrm{d}}{\mathrm{dx}}\left(\sum_{n=1}^{N_{E}}  a_{n}^{E}\, c_{E}^{-b_{n}^{E}}\,\gamma\left(b_{n}^{E}, \frac{\sigma_E^{2}}{P_E} x\right) \right), 
\end{align}
or equivalently
\begin{align}
     f_{\gamma_{E}}\left(x\right) = \sum_{n=1}^{N_{E}}  a_{n}^{E}\, c_{E}^{-b_{n}^{E}}\, \frac{\mathrm{d}}{\mathrm{dx}}\left(\gamma\left(b_{n}^{E}, \frac{\sigma_E^{2}}{P_E} x\right) \right).
     \label{Eq:f_gamma_E_step_3}
\end{align}
By applying~\cite[Eq. (8.350/1)]{B:Gra_Ryz_Book} in~\eqref{Eq:f_gamma_E_step_3}, we obtain~\eqref{Eq:f_gamma_E}. This concludes the proof.

\section*{Appendix C}
\section*{Proof of Theorem 1}
The probability of information leakage is defined~as
\begin{align}
    P_l = \Pr\left(R_s\leq 0\right),
\end{align}
which, based on~\eqref{Eq:Rs_instance}, can be rewritten~as
\begin{align}
    P_l = \Pr\left(C_B - C_E \leq 0 \right),
\end{align}
or equivalently
\begin{align}
    P_l = \Pr\left(C_B \leq C_E \right).
    \label{Eq:P_l_step_1}
\end{align}
By applying~\eqref{Eq:C_B} and~\eqref{Eq:C_E},~\eqref{Eq:P_l_step_1} can be expressed~as
\begin{align}
    P_l = \Pr\left(\gamma_B \leq \gamma_E \right).
    \label{Eq:P_l_step_2}
\end{align}
Since $\gamma_B$ and $\gamma_E$ are independent RVs,~\eqref{Eq:P_l_step_2} can be evaluated~as
\begin{align}
    P_l = \int_{0}^{\infty} F_{\gamma_B}\left(x\right)\,f_{\gamma_E}\left(x\right)\,\mathrm{d}x.
    \label{Eq:P_l_step_3}
\end{align}
With the aid of~\eqref{Eq:F_gamma_B} and~\eqref{Eq:f_gamma_E},~\eqref{Eq:P_l_step_3} can be written as 
\begin{align}
    P_l = \frac{1}{\Gamma\left(k_A\right)\Gamma\left(m_A\right)} \sum_{n=1}^{N_{E}}  a_{n}^{E}\,
    \left(\frac{\sigma_E^{2}}{P_E c_{E}}\right)^{b_{n}^{E}}\,  \mathcal{J}(n),
    \label{Eq:P_l_step_4}
\end{align}
where
\begin{align}
    \mathcal{J}(n)=\int_{0}^{\infty} & x^{b_{n}^{E}-1}\,\exp\left(-\frac{\sigma_E^{2}}{P_E} x \right)  
    \nonumber \\ \times &\G_{1, 3}^{2, 1}\left(\Xi^2 \frac{x\,\sigma_B^2}{P_B}\left|\begin{array}{c} 1\\ k_A, m_A, 0 \end{array}\right.\right)\,\mathrm{d}x.
    \label{Eq:J_n}
\end{align}
By applying~\cite{S:exp},~\eqref{Eq:J_n} can be written as
\begin{align}
    \mathcal{J}(n)=\int_{0}^{\infty} & x^{b_{n}^{E}-1}\,\G_{0,1}^{1,0}\left(\frac{\sigma_E^{2}}{P_E} x \left| 0 \right.\right) \nonumber \\ \times &\G_{1, 3}^{2, 1}\left(\Xi^2 \frac{x\,\sigma_B^2}{P_B}\left|\begin{array}{c} 1\\ k_A, m_A, 0 \end{array}\right.\right)\,\mathrm{d}x,
\end{align}
which, with the aid of~\cite{S:meijerG_int}, can be expressed in closed-form~as
\begin{align}
    \mathcal{J}(n)= \left(\frac{P_E}{\sigma_E^{2}}\right)^{b_n^{E}} 
    \G_{2,3}^{2,2}\left(\frac{\Xi^{2}\,\sigma_B^2\,P_{E}}{\sigma_{E}^{2}\,P_B}\left|\begin{array}{c} 1, 1-b_n^{E}\\ k_A, m_A, 0\end{array}\right.\right).
    \label{Eq:J_n_step_2}
\end{align}
With the aid of~\eqref{Eq:J_n_step_2},~\eqref{Eq:P_l_step_4} gives~\eqref{Eq:P_l}. This concludes the proof.

\section*{Appendix D}
\section*{Proof of Theorem 2}
For the case Alice have perfect and full CSI knowledge, the average secrecy rate of the RIS-empowered wireless system is defined~as
\begin{align}
    \tilde{C}_s = \mathbb{E}\left[C_s\right],
\end{align}
which, by applying~\eqref{Eq:C_s}, can be rewritten~as
\begin{align}
    \tilde{C}_s = \mathbb{E}\left[\max\left(C_B-C_E, 0\right)\right].
\end{align}
or equivalently as in~\eqref{Eq:av_C_s_1}, given at the top of the next page.
\begin{figure*}
\begin{align}
    \tilde{C}_s = \int_{0}^{\infty} \log_2\left(1+x_B\right)\,f_{\gamma_B}\left(x_B\right)\left( \int_{0}^{x_B}f_{\gamma_{E}}\left(x_{E}\right)\,\mathrm{d}x_E \right)\mathrm{d}x_B
    - \int_{0}^{\infty} \log_2\left(1+x_E\right)\,f_{\gamma_E}\left(x_E\right)\left( \int_{0}^{x_E}f_{\gamma_{B}}\left(x_{B}\right)\,\mathrm{d}x_B \right)\mathrm{d}x_E
    \label{Eq:av_C_s_1}
\end{align}
\hrulefill 
\end{figure*}
Notice that 
\begin{align}
    F_{\gamma_E}\left(x_B\right) = \int_{0}^{x_B}f_{\gamma_{E}}\left(x_{E}\right)\,\mathrm{d}x_E
\end{align}
and
\begin{align}
    F_{\gamma_B}\left(x_E\right) = \int_{0}^{x_E}f_{\gamma_{B}}\left(x_{B}\right)\,\mathrm{d}x_B;
\end{align}
thus,~\eqref{Eq:av_C_s_1} can be rewritten as in~\eqref{Eq:av_C_s_2}, given at the top of the next page. 
\begin{align}
    \tilde{C}_s =\mathcal{I}_1 - \mathcal{I}_2
    \label{Eq:av_C_s_2}
\end{align}
where
\begin{align}
    \mathcal{I}_1 = \int_{0}^{\infty} \log_2\left(1+x_B\right)\,f_{\gamma_B}\left(x_B\right) F_{\gamma_{E}}\left(x_{B}\right)\mathrm{d}x_B
    \label{Eq:I1}
\end{align}
and
\begin{align}
    \mathcal{I}_2 =\int_{0}^{\infty} \log_2\left(1+x_E\right)\,f_{\gamma_E}\left(x_E\right)F_{\gamma_{B}}\left(x_{E}\right)\mathrm{d}x_E.
    \label{Eq:I2}
\end{align}

By applying~\eqref{Eq:f_gamma_B} and~\eqref{Eq:F_gamma_E},~\eqref{Eq:I1} can be written~as in~\eqref{Eq:I1_step1}, given at the top of the next page. 
\begin{align}
    \mathcal{I}_1 = 
    \frac{1}{\Gamma\left(k_A\right)\Gamma\left(m_A\right)} \left(\frac{\Xi^2\,\sigma_B^2}{P_B} \right)^{\frac{k_A+m_A}{2}} \sum_{n=1}^{N_{E}} a_{n}^{E}\, c_{E}^{-b_{n}^{E}}\,
    \mathcal{I}_{1,1}(n),
    \label{Eq:I1_step1}
\end{align}
where 
\begin{align}
    \mathcal{I}_{1,1}(n)= \int_{0}^{\infty} & x^{\frac{k_A+m_A}{2}-1}\, \log_2\left(1+x_B\right)\, \gamma\left(b_{n}^{E}, \frac{\sigma_E^{2}}{P_E} x_B\right)\nonumber \\ & \times \G_{0,2}^{2,0}\left(\frac{\Xi^2\,\sigma_B^2}{2\,P_B} x_B \left| \frac{k_{A}-m_{A}}{2}, \frac{m_A-k_A}{2}\right.\right)  \mathrm{d}x_B.
    \label{Eq:I_11_def}
\end{align}
By applying~\cite{S:log},~\eqref{Eq:I_11_def} can be rewritten~as~\eqref{Eq:I_11_step1}, given at the top of the next page.
\begin{figure*}
\begin{align}
    \mathcal{I}_{1,1}(n)= \frac{1}{\ln(2)} \int_{0}^{\infty} & x^{\frac{k_A+m_A}{2}-1}\,  \gamma\left(b_{n}^{E}, \frac{\sigma_E^{2}}{P_E} x_B\right)
    \G_{2,2}^{1,2}\left(x_B\left|\begin{array}{l} 1, 1\\ 1, 0\end{array}\right.\right)
    \G_{0,2}^{2,0}\left(\frac{\Xi^2\,\sigma_B^2}{2\,P_B} x_B \left| \frac{k_{A}-m_{A}}{2}, \frac{m_A-k_A}{2}\right. \right)  \mathrm{d}x_B.
    \label{Eq:I_11_step1}
\end{align}
\hrulefill
\end{figure*}
Next, by employing~\cite{S:lower_gamma},~\eqref{Eq:I_11_step1} can be written as~\eqref{Eq:I_11_step2}, given at the top of the next page.
\begin{figure*}
\begin{align}
    \mathcal{I}_{1,1}(n)= \frac{1}{\ln(2)} \int_{0}^{\infty} & x^{\frac{k_A+m_A}{2}-1}\, \G_{1,2}^{1,1}\left( \frac{\sigma_E^{2}}{P_E} x_B \left|\begin{array}{c} 1 \\ b_{n}^{E}, 0 \end{array}\right.\right)
    \G_{2,2}^{1,2}\left(x_B\left|\begin{array}{l} 1, 1\\ 1, 0\end{array}\right.\right)
    \G_{0,2}^{2,0}\left(\frac{\Xi^2\,\sigma_B^2}{2\,P_B} x_B \left| \frac{k_{A}-m_{A}}{2}, \frac{m_A-k_A}{2}\right. \right)  \mathrm{d}x_B.
    \label{Eq:I_11_step2}
\end{align}
\hrulefill
\end{figure*} 
Finally, by applying~\cite{S:genMeijer}, we obtain~\eqref{Eq:I_11_step3}, given at the top of the next page.
\begin{figure*}
\begin{align}
    \mathcal{I}_{1,1}(n)= \frac{1}{\ln(2)} \left(\frac{\sigma_E^2}{P_E}\right)^{-\frac{k_A+m_A}{2}} 
    \G_{1,2:2,2:0,0}^{1,1:1,2:2,0}\left( 
        \begin{array}{c}  
            1-\frac{k_A+m_A}{2}-b_n^{E}, 1-\frac{k_A+m_A}{2} \\ -\frac{k_A+m_A}{2}
        \end{array}
        \left| 
         \begin{array}{c}  
            1, 1 \\ 1, 0
        \end{array}
        \right.
        \left| 
        \begin{array}{c}  
             - \\
             \frac{k_A-m_A}{2}, \frac{m_A-k_A}{2}
        \end{array}
        \right|
        \frac{P_E}{\sigma_E^2}, \frac{\Xi^2\,\sigma_{B}^{2}\,P_E}{2\,P_B\,\sigma_{E}^{2}}
    \right)
    \label{Eq:I_11_step3}
\end{align}
\hrulefill
\end{figure*} 
By applying~\eqref{Eq:I_11_step3} in~\eqref{Eq:I1_step1}, we obtain~\eqref{Eq:I1_step2}, given at the top of the next page. 
\begin{figure*}
    \begin{align}
    \mathcal{I}_1 &= \frac{1}{\ln(2)}
    \frac{1}{\Gamma\left(k_A\right)\Gamma\left(m_A\right)} 
    \left(\frac{\Xi^2\, P_E\,\sigma_B^2}{P_B\,\sigma_E^2} \right)^{\frac{k_A+m_A}{2}} 
    \nonumber \\ & \times
    \sum_{n=1}^{N_{E}} a_{n}^{E}\, c_{E}^{-b_{n}^{E}}\, 
    \G_{1,2:2,2:0,0}^{1,1:1,2:2,0}\left( 
        \begin{array}{c}  
            1-\frac{k_A+m_A}{2}-b_n^{E}, 1-\frac{k_A+m_A}{2} \\ -\frac{k_A+m_A}{2}
        \end{array}
        \left| 
         \begin{array}{c}  
            1, 1 \\ 1, 0
        \end{array}
        \right.
        \left| 
        \begin{array}{c}  
             - \\
             \frac{k_A-m_A}{2}, \frac{m_A-k_A}{2}
        \end{array}
        \right|
        \frac{P_E}{\sigma_E^2}, \frac{\Xi^2\,\sigma_{B}^{2}\,P_E}{2\,P_B\,\sigma_{E}^{2}}
    \right)
    \label{Eq:I1_step2}
\end{align}
\hrulefill
\end{figure*}

Next, we use~\eqref{Eq:f_gamma_E} and~\eqref{Eq:F_gamma_B} in order to rewrite~\eqref{Eq:I2}~as
\begin{align}
    \mathcal{I}_2 =\frac{1}{\Gamma\left(k_A\right)\Gamma\left(m_A\right)}\sum_{n=1}^{N_{E}}  a_{n}^{E}\,
    \left(\frac{\sigma_E^{2}}{P_E c_{E}}\right)^{b_{n}^{E}}\, \mathcal{I}_{2,1}(n),
    \label{Eq:I2step1}
\end{align}
where
\begin{align}
    \mathcal{I}_{2,1}(n)=&\int_{0}^{\infty} 
    x_E^{b_{n}^{E}-1}\,\exp\left(-\frac{\sigma_E^{2}}{P_E} x_E \right)
    \log_2\left(1+x_E\right)
    \nonumber \\ & \times
    \G_{1, 3}^{2, 1}\left(\Xi^2 \frac{x_E\,\sigma_B^2}{P_B}\left|\begin{array}{c} 1\\ k_A, m_A, 0 \end{array}\right.\right)\mathrm{d}x_E.
    \label{Eq:I21_step1}
\end{align}
With the aid of~\cite{S:log},~\eqref{Eq:I21_step1} can be rewritten~as
\begin{align}
    \mathcal{I}_{2,1}(n)=&\frac{1}{\ln(2)}\int_{0}^{\infty} 
    x_E^{b_{n}^{E}-1}\,\exp\left(-\frac{\sigma_E^{2}}{P_E} x_E \right)
    \G_{2,2}^{1,2}\left(x_{E}\left|\begin{array}{c}1, 1 \\ 1, 0\end{array}\right.\right)
    \nonumber \\ & \times
    \G_{1, 3}^{2, 1}\left(\Xi^2 \frac{x_E\,\sigma_B^2}{P_B}\left|\begin{array}{c} 1\\ k_A, m_A, 0 \end{array}\right.\right)\mathrm{d}x_E,
    \label{Eq:I21_step2}
\end{align}
which, after applying~\cite{S:exp}, can be expressed~as
\begin{align}
    &\mathcal{I}_{2,1}(n)=\frac{1}{\ln(2)}\int_{0}^{\infty} 
    x_E^{b_{n}^{E}-1}\,\G_{0,1}^{1,0}\left(\frac{\sigma_E^{2}}{P_E} x_E\left| 0\right. \right)
    \nonumber \\ & \times
    \G_{2,2}^{1,2}\left(x_{E}\left|\begin{array}{c}1, 1 \\ 1, 0\end{array}\right.\right)
    \G_{1, 3}^{2, 1}\left(\Xi^2 \frac{x_E\,\sigma_B^2}{P_B}\left|\begin{array}{c} 1\\ k_A, m_A, 0 \end{array}\right.\right)\mathrm{d}x_E
    \label{Eq:I21_step3}
\end{align}
Next, by applying~\cite{S:genMeijer}, we obtain~\eqref{Eq:I21_step4}, given at the top of the next page.
\begin{figure*}
\begin{align}
    &\mathcal{I}_{2,1}(n)=\frac{1}{\ln(2)} \G_{1,0:2,2:1,3}^{0,1:1,2:2,1}\left( 1-b_n^{E} \left|\begin{array}{c} 1, 1 \\ 1, 0 \end{array}\right|\left.\begin{array}{c} 1\\ k_A, m_A, 0 \end{array}\right| \frac{P_E}{\sigma_E^2}, \frac{\Xi^2 \sigma_B^2 P_E}{\sigma_E^2\,P_B} \right)
    \label{Eq:I21_step4}
\end{align}
\hrulefill
\end{figure*}

With the aid of~\eqref{Eq:I21_step4}, \eqref{Eq:I2step1} can be written as~\eqref{Eq:I2}, given at the top of the next page. 
\begin{figure*}
\begin{align}
    \mathcal{I}_2=\frac{1}{\ln(2)}\frac{1}{\Gamma\left(k_A\right)\Gamma\left(m_A\right)}\sum_{n=1}^{N_{E}}  a_{n}^{E}\,
    \left(\frac{\sigma_E^{2}}{P_E c_{E}}\right)^{b_{n}^{E}}\, \G_{1,0:2,2:1,3}^{0,1:1,2:2,1}\left( 1-b_n^{E} \left|\begin{array}{c} 1, 1 \\ 1, 0 \end{array}\right|\left.\begin{array}{c} 1\\ k_A, m_A, 0 \end{array}\right| \frac{P_E}{\sigma_E^2}, \frac{\Xi^2 \sigma_B^2 P_E}{\sigma_E^2\,P_B} \right)
    \label{Eq:I2}
\end{align}
\hrulefill
\end{figure*}
Finally, by applying~\eqref{Eq:I1} and~\eqref{Eq:I2} to~\eqref{Eq:av_C_s_2}, we obtain~\eqref{Eq:av_Cs_final}. This concludes the proof.

\section*{Appendix E}
\section*{Proof of Theorem 3}

For the case Alice has no knowledge of the instantaneous characteristics of the Alice-Eve channel, the average secrecy rate is defined~as
\begin{align}
    \tilde{R}_{s} = \mathbb{E}\left[R_s\right].
    \label{Eq:av_R_s_def}
\end{align}
By applying~\eqref{Eq:Rs_instance},~\eqref{Eq:av_R_s_def} can be expressed~as
\begin{align}
    \tilde{R}_{s} = \mathbb{E}\left[C_B-C_E\right],
\end{align}
or
\begin{align}
    \tilde{R}_s = \tilde{C}_B - \tilde{C}_E,
    \label{Eq:R_s_def}
\end{align}
where 
\begin{align}
    \tilde{C}_B = \mathbb{E}\left[C_B\right]
\end{align}
and
\begin{align}
    \tilde{C}_E = \mathbb{E}\left[C_E\right],
\end{align}
are the ergodic capacities of the legitimate and eavesdropping channel, respectively. 

The ergodic capacity of the legitimate channel can be evaluated~as
\begin{align}
	\overline{C}_B = \int_{0}^{\infty} \log_2\left(1+\frac{P_B}{\sigma_{B}^{2}} x^2\right)\, f_{A_r}(x)\,\mathrm{d}x,
\end{align}
which, with the aid of~\eqref{Eq:f_A}, can be written~as
\begin{align}
\overline{C}_B = \frac{4 \Xi^{k_A + m_A}}{\Gamma\left(k_A\right)\Gamma\left(m_A\right)} \mathcal{K},
\label{Eq:C_B_step_2}
\end{align}
where
\begin{align}
	\mathcal{K}= \int_{0}^{\infty} x^{k_A+m_A-1} \log_2\left(1+\frac{P_B}{\sigma_B^2} x^2\right)  \mathrm{K}_{k_A-m_A}\left(2\Xi x\right)\,\mathrm{d}x.
	\label{Eq:K}
\end{align}
To derive a closed-form expression for~\eqref{Eq:K}, we first use~\cite[Eq. (15.1.1)]{B:Abramowitz}, and express~\eqref{Eq:K}~as
\begin{align}
		\mathcal{K}= \frac{P_B}{\ln(2)\, \sigma_B^{2}} \int_{0}^{\infty} & x^{k_A+m_A+1} \,_2F_1\left(1, 1; 2; -\frac{P_B}{\sigma_B^{2}} x^2\right) 
		\nonumber \\ & \times 
		\mathrm{K}_{k_A-m_A}\left(2\Xi x\right)\,\mathrm{d}x,
\end{align}
which, with the aid of~\cite[Eq. (9.34/7)]{B:Gra_Ryz_Book}, can be rewritten~as
\begin{align}
	 \mathcal{K}= \frac{P_B}{\ln(2)\, \sigma_B^{2}} \int_{0}^{\infty} & x^{k_A+m_A+1} \, \mathrm{G}_{2, 2}^{1, 2}\left( \frac{P_B}{\sigma_B^{2}}\,x^2\left|\begin{array}{c} 0, 0 \\ 0, -1\end{array}\right.\right)
	\nonumber \\ & \times 
	\mathrm{K}_{k_A-m_A}\left(2\Xi x\right)\,\mathrm{d}x.
    \label{Eq:K_step2}
\end{align}
By applying~\cite{Wofram:BesselK} to~\eqref{Eq:K_step2}, we obtain
\begin{align}
	\mathcal{K}= \frac{P_B}{2\ln(2)\, \sigma_B^{2}} &\int_{0}^{\infty}  x^{k_A+m_A+1} \, \mathrm{G}_{2, 2}^{1, 2}\left( \frac{P_B}{\sigma_B^{2}}\,x^2\left|\begin{array}{c} 0, 0 \\ 0, -1\end{array}\right.\right)
	\nonumber \\ & \times 
	\mathrm{G}_{0,2}^{2,0}\left({\Xi^2\,x^2}\left|\begin{array}{c} \frac{k_A-m_A}{2}, - \frac{k_A-m_A}{2}\end{array}\right.\right)
	\,\mathrm{d}x.
	\label{Eq:K_step3}
\end{align}
Next, we perform a variable change in~\eqref{Eq:K_step3} by setting $y=x^2$, and we get
\begin{align}
	\mathcal{K}= &\frac{P_B}{4\ln(2)\, \sigma_B^{2}} \int_{0}^{\infty}  y^{\frac{1}{2}\left(k_A+m_A\right)} \, \mathrm{G}_{2, 2}^{1, 2}\left( \frac{P_B}{\sigma_B^{2}}\,y\left|\begin{array}{c} 0, 0 \\ 0, -1\end{array}\right.\right)
	\nonumber \\ & \times 
	\mathrm{G}_{0,2}^{2,0}\left({\Xi^2}\,y\left|\begin{array}{c} \frac{k_A-m_A}{2}, - \frac{k_A-m_A}{2}\end{array}\right.\right)
	\,\mathrm{d}y.
	\label{Eq:K_step4}
\end{align}
By applying~\cite[ch. 2.3]{Mathaia2010},~\eqref{Eq:K_step4} can be expressed~as
\begin{align}
		&\mathcal{K}= \frac{1}{4\ln(2)}\left(\frac{P_B}{\sigma_B^{2}}\right)^{-\frac{k_A+m_A}{2}}
		\nonumber \\ & \times
		\mathrm{G}_{2,4}^{4,1}\left(\frac{\Xi^2\,\sigma_B^{2}}{P_B}\left|\begin{array}{c} -\frac{k_A+m_A}{2}, 1-\frac{k_A+m_A}{2}\\ \frac{k_A-m_A}{2}, - \frac{k_A-m_A}{2}, -\frac{k_A+m_A}{2}, -\frac{k_A+m_A}{2}\end{array}\right.\right).
		\label{Eq:K_step5}
\end{align}
From~\eqref{Eq:K_step5},~\eqref{Eq:C_B_step_2} can be expressed in closed-form~as 
\begin{align}
	&\overline{C}_B =\frac{1}{\ln(2)}\left(\frac{P_B}{\sigma_B^{2}}\right)^{-\frac{k_A+m_A}{2}}  \frac{ \Xi^{k_A + m_A}}{\Gamma\left(k_A\right)\Gamma\left(m_A\right)}
	\nonumber \\ & \times  \mathrm{G}_{2,4}^{4,1}\left(\frac{\Xi^2\,\sigma_B^{2}}{P_B}\left|\begin{array}{c} -\frac{k_A+m_A}{2}, 1-\frac{k_A+m_A}{2}\\ \frac{k_A-m_A}{2}, - \frac{k_A-m_A}{2}, -\frac{k_A+m_A}{2}, -\frac{k_A+m_A}{2}\end{array}\right.\right).
	\label{Eq:C_B_final}
\end{align}

The ergodic capacity of the eavesdropping channel can be evaluated~as 
\begin{align}
	\overline{C}_E = \int_{0}^{\infty} \log_{2}\left(1+\frac{P_{E}}{\sigma_E^{2}} x^2\right) \, f_{h_{E}}\left(x\right)\,\mathrm{d}x.
    \label{Eq:C_E_def}
\end{align}
With the aid of~\eqref{Eq:f_h_E},~\eqref{Eq:C_E_def} can be expressed~as 
\begin{align}
	\overline{C}_E = \sum_{n=1}^{N_E} 2\,a_{n}^{E}\, \mathcal{L}(n),
	\label{Eq:C_E_step_3}
\end{align}
where 
\begin{align}
	\mathcal{L}(n) =  \int_{0}^{\infty} x^{2\,b_n^{E}-1} \, \exp\left(-c_E\,x^2\right)\, \log_{2}\left(1+\frac{P_{B}}{\sigma_B^{2}} x^2\right) \,\mathrm{d}x.
	\label{Eq:L_n_def}
\end{align}
By applying basic logarithmic properties,~\eqref{Eq:L_n_def} can be expressed~as
\begin{align}
	\mathcal{L}(n) = \frac{1}{\ln(2)} \int_{0}^{\infty} &x^{2\,b_n^{E}-1} \, \exp\left(-c_E\,x^2\right)\,
	\nonumber \\ & \times \ln\left(1+\frac{P_{E}}{\sigma_E^{2}} x^2\right) \,\mathrm{d}x.
	\label{Eq:L_n_step_1}
\end{align}
Next, we set $z=x^{2}$ in~\eqref{Eq:L_n_step_1} and we obtain
 \begin{align}
 	\mathcal{L}(n) = \frac{1}{2\ln(2)} \int_{0}^{\infty} &z^{\,b_n^{E}-1} \, \exp\left(-c_E\,z\right)\,
 	\nonumber \\ & \times \ln\left(1+\frac{P_{E}}{\sigma_{E}^{2}} z\right) \,\mathrm{d}z.
 	\label{Eq:L_n_step_2}
 \end{align}
With the aid of~\cite[Eq. (8.352/2)]{B:Gra_Ryz_Book},~\eqref{Eq:L_n_step_2} can be written~as
\begin{align}
    \mathcal{L}(n) = \frac{1}{2\ln(2)} \int_{0}^{\infty} &z^{\,b_n^{E}-1} \, \exp\left(-c_E\,z\right)\,
	\nonumber \\ & \times
	\mathrm{G}_{2, 2}^{1, 2}\left( \frac{P_E}{\sigma_E^{2}}\,z\left|\begin{array}{c} 0, 0 \\ 0, -1\end{array}\right.\right) \,\mathrm{d}z,
	\label{Eq:L_n_step_3}
\end{align}
which, by applying~\cite[Eq. (8.352/2)]{B:Gra_Ryz_Book}, gives
 \begin{align}
	\mathcal{L}(n) = \frac{1}{2\ln(2)} \int_{0}^{\infty} &z^{\,b_n^{E}-1} \, \Gamma\left(1, c_E\,z\right)\,
	\nonumber \\ & \times
	\mathrm{G}_{2, 2}^{1, 2}\left( \frac{P_E}{\sigma_E^{2}}\,z\left|\begin{array}{c} 0, 0 \\ 0, -1\end{array}\right.\right) \,\mathrm{d}z.
	\label{Eq:L_n_step_4}
\end{align}
With the use of~\cite[Eq. (15.1.1)]{B:Abramowitz} and~\cite[Eq. (9.34/7)]{B:Gra_Ryz_Book},~\eqref{Eq:L_n_step_4}~yields
\begin{align}
	\mathcal{L}(n) = \frac{1}{2\ln(2)} \int_{0}^{\infty} &z^{\,b_n^{E}-1} \, 
	\mathrm{G}_{1,2}^{2,0}\left(c_E\,z\left|\begin{array}{c} 1\\0, 1\end{array}\right.\right)
	\nonumber \\ & \times
	\mathrm{G}_{2, 2}^{1, 2}\left( \frac{P_E}{\sigma_E^{2}}\,z\left|\begin{array}{c} 0, 0 \\ 0, -1\end{array}\right.\right) \,\mathrm{d}z,
	\label{Eq:L_n_step_5}
\end{align}
which, by applying~\cite[ch. 2.3]{Mathaia2010}, returns
\begin{align}
	\mathcal{L}(n) = \frac{c_E^{-b_n^{E}}}{2\ln(2)} \mathrm{G}_{4,3}^{1,4}\left(\frac{P_E}{c_E\, \sigma_E^{2}}\left|\begin{array}{c} 0, 0, 1-b_n^{E}, -b_n^{E}\\ 0, -b_n^{E}, -1 \end{array}\right.\right).
	\label{Eq:L_n_step_6}
\end{align}
By applying~\eqref{Eq:L_n_step_6} in~\eqref{Eq:C_E_step_3}, we~obtain 
\begin{align}
	\overline{C}_E = \sum_{n=1}^{N_{E}} & \frac{ a_{n}^{E}\, c_E^{-b_n^{E}}}{\ln(2)} 
	\nonumber \\ & \times \mathrm{G}_{4,3}^{1,4}\left(\frac{P_E}{c_E\, \sigma_E^{2}}\left|\begin{array}{c} 0, 0, 1-b_n^{E}, -b_n^{E}\\ 0, -b_n^{E}, -1 \end{array}\right.\right).
	\label{Eq:C_E_step_final}
\end{align}
Finally, with the aid of~\eqref{Eq:C_B_final} and~\eqref{Eq:C_E_step_final},~\eqref{Eq:R_s_def} can be written as in~\eqref{Eq:R_s_final}. This concludes the proof.

\bibliographystyle{IEEEtran}
\bibliography{IEEEabrv,References}

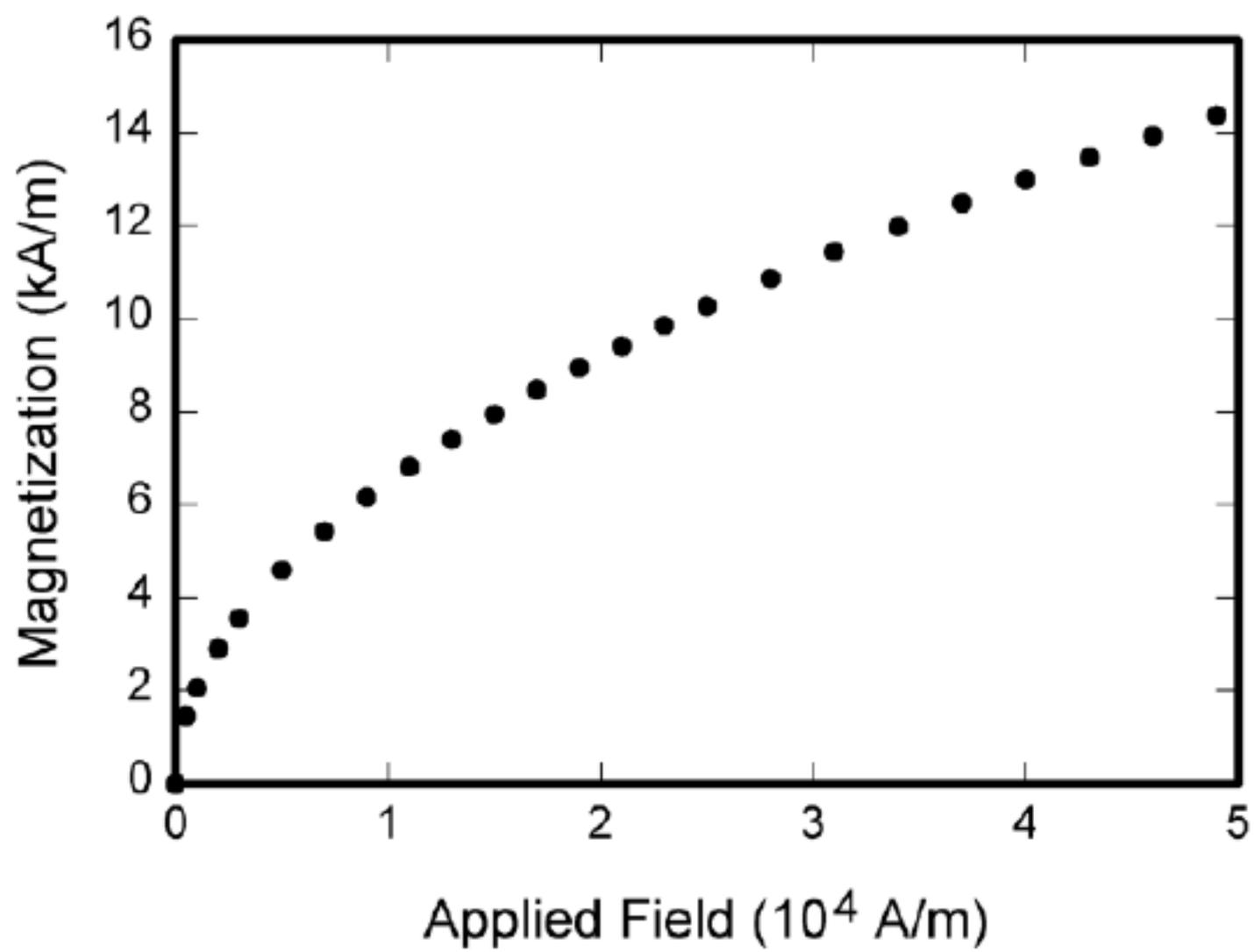
\end{document}